# Properties of the Binary Black Hole Merger GW150914

B. P. Abbott *et al.**

(LIGO Scientific Collaboration and Virgo Collaboration)

On September 14, 2015, the Laser Interferometer Gravitational-Wave Observatory (LIGO) detected a gravitational-wave transient (GW150914); we characterize the properties of the source and its parameters. The data around the time of the event were analyzed coherently across the LIGO network using a suite of accurate waveform models that describe gravitational waves from a compact binary system in general relativity. GW150914 was produced by a nearly equal mass binary black hole of masses $36^{+5}_{-4} M_\odot$ and $29^{+4}_{-4} M_\odot$; for each parameter we report the median value and the range of the 90% credible interval. The dimensionless spin magnitude of the more massive black hole is bound to be $< 0.7$ (at 90% probability). The luminosity distance to the source is $410^{+160}_{-180}$ Mpc, corresponding to a redshift $0.09^{+0.03}_{-0.04}$ assuming standard cosmology. The source location is constrained to an annulus section of 610 deg$^2$, primarily in the southern hemisphere. The binary merges into a black hole of mass $62^{+4}_{-4} M_\odot$ and spin $0.67^{+0.05}_{-0.07}$. This black hole is significantly more massive than any other inferred from electromagnetic observations in the stellar-mass regime.

## I. INTRODUCTION

In Ref. [1] we reported the detection of gravitational waves (GWs), observed on September 14, 2015 at 09:50:45 UTC, by the twin instruments of the twin instruments of the Laser Interferometer Gravitational-Wave Observatory (LIGO) located at Hanford, Washington, and Livingston, Louisiana, in the USA [2,3]. The transient signal, named GW150914, was detected with a false-alarm-probability of $< 2 \times 10^{-7}$ and has been associated with the merger of a binary system of black holes (BHs).

Here we discuss the properties of this source and its inferred parameters. The results are based on a complete analysis of the data surrounding this event. The only information from the search stage is the time of arrival of the signal. Crucially, this analysis differs from the search in the following fundamental ways: it is coherent across the LIGO network, it uses waveform models that include the full richness of the physics introduced by BH spins, and it covers the full multidimensional parameter space of the considered models with a fine (stochastic) sampling; we also account for uncertainty in the calibration of the measured strain. The results of this analysis provide the parameter values quoted in Ref. [1] describing the properties of the source of GW150914. (Following the publication of Ref. [1] we repeated the analysis after fixing minor errors in the coordinate transformation between precessing and nonprecessing binary systems that affects waveforms used in the analysis. The parameter values reported here, in Table I and the figures, are the updated values and should be used for future studies. The only value different from what was reported in Ref. [1] is the total energy radiated, which previously was $3.0^{+0.5}_{-0.5} M_\odot c^2$ and now rounds to $3.0^{+0.5}_{-0.4} M_\odot c^2$).

In general relativity, two objects in orbit slowly spiral together due to the loss of energy and angular momentum through gravitational radiation [7,8]. This is in contrast to Newtonian gravity where bodies can follow closed, elliptical orbits [9,10]. As the binary shrinks, the frequency and amplitude of the emitted GWs increase. Eventually the two objects merge. If these objects are BHs, they form a single perturbed BH that radiates GWs as a superposition of quasinormal ringdown modes. Typically, one mode dominates soon after merger, and an exponentially damped oscillation at constant frequency can be observed as the BH settles to its final state [11,12].

An isolated BH is described by only its mass and spin, since we expect the electric charge of astrophysical BHs to be negligible [13–16]. Merging (BBHs) are therefore relatively simple systems. The two BHs are described by eight intrinsic parameters: the masses $m_{1,2}$ (defined as the gravitational masses of the BHs in isolation) and spins $\boldsymbol{S}_{1,2}$ (magnitude and orientation) of the individual BHs. For a BH of mass $m$, the spin can be at most $Gm^2/c$; hence, it is conventional to quote the dimensionless spin magnitude $a = c|\boldsymbol{S}|/(Gm^2) \leq 1$. Nine additional parameters are needed to fully describe the binary: the location (luminosity distance $D_L$, right ascension $\alpha$, and declination $\delta$); orientation (the binary's orbital inclination $\iota$ and polarization $\psi$);

*Full author list given at the end of the article.

TABLE I. Summary of the parameters that characterize GW150914. For model parameters we report the median value as well as the range of the symmetric 90% credible interval [4]; where useful, we also quote 90% credible bounds. For the logarithm of the Bayes factor for a signal compared to Gaussian noise we report the mean and its 90% standard error from 4 parallel runs with a nested sampling algorithm [5]. The source redshift and source-frame masses assume standard cosmology [6]. The spin-aligned EOBNR and precessing IMRPhenom waveform models are described in the text. Results for the effective precession spin parameter $\chi_p$ used in the IMRPhenom model are not shown as we effectively recover the prior; we constrain $\chi_p < 0.71$ at 90% probability, see left panel of Fig. 5. The Overall results are computed by averaging the posteriors for the two models. For the Overall results we quote both the 90% credible interval or bound and an estimate for the 90% range of systematic error on this determined from the variance between waveform models. The sky location associated with GW150914 is presented in Fig. 4 and discussed in the text.

| | EOBNR | IMRPhenom | Overall |
|---|---|---|---|
| Detector-frame total mass $M/M_\odot$ | $70.3^{+5.3}_{-4.8}$ | $70.9^{+4.0}_{-3.9}$ | $70.6^{+4.6\pm0.5}_{-4.5\pm1.3}$ |
| Detector-frame chirp mass $\mathcal{M}/M_\odot$ | $30.2^{+2.5}_{-1.9}$ | $30.6^{+1.8}_{-1.8}$ | $30.4^{+2.1\pm0.2}_{-1.9\pm0.5}$ |
| Detector-frame primary mass $m_1/M_\odot$ | $39.4^{+5.5}_{-4.9}$ | $38.5^{+5.6}_{-3.6}$ | $38.9^{+5.6\pm0.6}_{-4.3\pm0.4}$ |
| Detector-frame secondary mass $m_2/M_\odot$ | $30.9^{+4.8}_{-4.4}$ | $32.2^{+3.6}_{-4.8}$ | $31.6^{+4.2\pm0.1}_{-4.7\pm0.9}$ |
| Detector-frame final mass $M_\mathrm{f}/M_\odot$ | $67.1^{+4.6}_{-4.4}$ | $67.6^{+3.6}_{-3.5}$ | $67.4^{+4.1\pm0.4}_{-4.0\pm1.2}$ |
| Source-frame total mass $M^\mathrm{source}/M_\odot$ | $65.0^{+5.0}_{-4.4}$ | $65.0^{+4.0}_{-3.6}$ | $65.0^{+4.5\pm0.8}_{-4.0\pm0.7}$ |
| Source-frame chirp mass $\mathcal{M}^\mathrm{source}/M_\odot$ | $27.9^{+2.3}_{-1.8}$ | $28.1^{+1.7}_{-1.6}$ | $28.0^{+2.0\pm0.3}_{-1.7\pm0.3}$ |
| Source-frame primary mass $m_1^\mathrm{source}/M_\odot$ | $36.3^{+5.3}_{-4.5}$ | $35.3^{+5.2}_{-3.4}$ | $35.8^{+5.3\pm0.9}_{-3.9\pm0.1}$ |
| Source-frame secondary mass $m_2^\mathrm{source}/M_\odot$ | $28.6^{+4.4}_{-4.2}$ | $29.6^{+3.3}_{-4.3}$ | $29.1^{+3.8\pm0.1}_{-4.3\pm0.7}$ |
| Source-frame final mass $M_\mathrm{f}^\mathrm{source}/M_\odot$ | $62.0^{+4.4}_{-4.0}$ | $62.0^{+3.7}_{-3.3}$ | $62.0^{+4.1\pm0.7}_{-3.7\pm0.6}$ |
| Mass ratio $q$ | $0.79^{+0.18}_{-0.19}$ | $0.84^{+0.14}_{-0.20}$ | $0.82^{+0.17\pm0.01}_{-0.20\pm0.03}$ |
| Effective inspiral spin parameter $\chi_\mathrm{eff}$ | $-0.09^{+0.19}_{-0.17}$ | $-0.05^{+0.13}_{-0.15}$ | $-0.07^{+0.16\pm0.01}_{-0.17\pm0.05}$ |
| Dimensionless primary spin magnitude $a_1$ | $0.32^{+0.45}_{-0.28}$ | $0.32^{+0.53}_{-0.29}$ | $0.32^{+0.49\pm0.06}_{-0.29\pm0.01}$ |
| Dimensionless secondary spin magnitude $a_2$ | $0.57^{+0.40}_{-0.51}$ | $0.34^{+0.54}_{-0.31}$ | $0.44^{+0.50\pm0.08}_{-0.40\pm0.02}$ |
| Final spin $a_\mathrm{f}$ | $0.67^{+0.06}_{-0.08}$ | $0.66^{+0.04}_{-0.06}$ | $0.67^{+0.05\pm0.01}_{-0.07\pm0.02}$ |
| Luminosity distance $D_L/\mathrm{Mpc}$ | $390^{+170}_{-180}$ | $440^{+150}_{-180}$ | $410^{+160\pm20}_{-180\pm40}$ |
| Source redshift $z$ | $0.083^{+0.033}_{-0.036}$ | $0.093^{+0.029}_{-0.036}$ | $0.088^{+0.032\pm0.005}_{-0.037\pm0.008}$ |
| Upper bound on primary spin magnitude $a_1$ | 0.65 | 0.74 | $0.69 \pm 0.08$ |
| Upper bound on secondary spin magnitude $a_2$ | 0.93 | 0.78 | $0.89 \pm 0.13$ |
| Lower bound on mass ratio $q$ | 0.64 | 0.68 | $0.66 \pm 0.03$ |
| Log Bayes factor $\ln \mathcal{B}_{s/n}$ | $288.7 \pm 0.2$ | $290.3 \pm 0.1$ | ... |

time $t_c$ and phase $\phi_c$ of coalescence, and two parameters describing eccentricity (the magnitude $e$ and the argument of periapsis) of the system.

Radiation reaction is efficient in circularizing orbits [17] before the signal enters the sensitive frequency band of the instruments (~20–1000 Hz). In our analysis, we assume circular orbits (we therefore do not include the eccentricity parameters), and we find no evidence for residual eccentricity, see the Discussion. Under the approximation of a circular orbit, the binary emits GWs primarily at twice the orbital frequency [18].

The gravitational waveform observed for GW150914 comprises ~10 cycles during the inspiral phase from 30 Hz, followed by the merger and ringdown. The properties of the binary affect the phase and amplitude evolution of the observed GWs, allowing us to measure the source parameters. Here we briefly summarize these signatures, and provide an insight into our ability to characterize the properties of GW150914 before we present the details of the Results; for methodological studies, we refer the reader to Refs. [4,19–24] and references therein.

In general relativity, gravitational radiation is fully described by two independent, and time-dependent polarizations, $h_+$ and $h_\times$. Each instrument $k$ measures the strain

$$h_k = F_k^{(+)} h_+ + F_k^{(\times)} h_\times, \tag{1}$$

a linear combination of the polarizations weighted by the antenna beam patterns $F_k^{(+,\times)}(\alpha, \delta, \psi)$, which depend on the source location in the sky and the polarization of the waves [25,26]. During the inspiral and at the leading order, the GW polarizations can be expressed as

$$h_+(t) = A_\mathrm{GW}(t)(1 + \cos^2\iota)\cos\phi_\mathrm{GW}(t), \tag{2a}$$

$$h_\times(t) = -2A_\mathrm{GW}(t)\cos\iota\sin\phi_\mathrm{GW}(t), \tag{2b}$$

where $A_\mathrm{GW}(t)$ and $\phi_\mathrm{GW}(t)$ are the GW amplitude and phase, respectively. For a binary viewed face-on ($\cos\iota = \pm1$),

GWs are circularly polarized, whereas for a binary observed edge-on ($\cos \iota = 0$), GWs are linearly polarized.

During the inspiral, the phase evolution $\phi_{\mathrm{GW}}(t; m_{1,2}, \boldsymbol{S}_{1,2})$ can be computed using post-Newtonian (PN) theory, which is a perturbative expansion in powers of the orbital velocity $v/c$ [27]. For GW150914, $v/c$ is in the range ≈0.2–0.5 in the LIGO sensitivity band. At the leading order, the phase evolution is driven by a particular combination of the two masses, commonly called the chirp mass [28],

$$\mathcal{M} = \frac{(m_1 m_2)^{3/5}}{M^{1/5}} \simeq \frac{c^3}{G}\left[\frac{5}{96}\pi^{-8/3} f^{-11/3} \dot{f}\right]^{3/5}, \qquad (3)$$

where $f$ is the GW frequency, $\dot{f}$ is its time derivative, and $M = m_1 + m_2$ is the total mass. Additional parameters enter at each of the following PN orders. First, the mass ratio, $q = m_2/m_1 \leq 1$, and the BH spin components parallel to the orbital angular momentum vector $\boldsymbol{L}$ affect the phase evolution. The full degrees of freedom of the spins enter at higher orders. Thus, from the inspiral, we expect to measure the chirp mass with highest accuracy and only place weak constraints on the mass ratio and (the components parallel to $\boldsymbol{L}$ of) the spins of the BHs [21,29].

Spins are responsible for an additional characteristic effect: if misaligned with respect to $\boldsymbol{L}$, they cause the binary's orbital plane to precess around the almost-constant direction of the total angular momentum of the binary, $\boldsymbol{J} = \boldsymbol{L} + \boldsymbol{S}_1 + \boldsymbol{S}_2$. This leaves characteristic amplitude and phase modulations in the observed strain [30,31], as $\psi$ and $\iota$ become time dependent. The size of these modulations depends crucially on the viewing angle of the source.

As the BHs get closer to each other and their velocities increase, the accuracy of the PN expansion degrades, and eventually the full solution of Einstein's equations is needed to accurately describe the binary evolution. This is accomplished using numerical relativity (NR) which, after the initial breakthrough [32–34], has been improved continuously to achieve the sophistication of modeling needed for our purposes. The details of the ringdown are primarily governed by the mass and spin of the final BH. In particular, the final mass and spin determine the (constant) frequency and decay time of the BH's ringdown to its final state [35]. The late stage of the coalescence allows us to measure the total mass which, combined with the measurement of the chirp mass and mass ratio from the early inspiral, yields estimates of the individual component masses for the binary.

The observed frequency of the signal is redshifted by a factor of $(1 + z)$, where $z$ is the cosmological redshift. There is no intrinsic mass or length scale in vacuum general relativity, and the dimensionless quantity that incorporates frequency is $fGm/c^3$. Consequently, a redshifting of frequency is indistinguishable from a rescaling of the masses by the same factor [20,36,37]. We therefore measure redshifted masses $m$, which are related to source frame masses by $m = (1 + z)m^{\mathrm{source}}$. However, the GW amplitude $A_{\mathrm{GW}}$, Eq. (2), also scales linearly with the mass and is inversely proportional to the comoving distance in an expanding universe. Therefore, the amplitude scales inversely with the luminosity distance, $A_{\mathrm{GW}} \propto 1/D_L$, and from the GW signal alone we can directly measure the luminosity distance, but not the redshift.

The observed time delay, and the need for the registered signal at the two sites to be consistent in amplitude and phase, allow us to localize the source to a ring on the sky [38,39]. Where there is no precession, changing the viewing angle of the system simply changes the observed waveform by an overall amplitude and phase. Furthermore, the two polarizations are the same up to overall amplitude and phase. Thus, for systems with minimal precession, the distance, binary orientation, phase at coalescence, and sky location of the source change the overall amplitude and phase of the source in each detector, but they do not change the signal morphology. Phase and amplitude consistency allow us to untangle some of the geometry of the source. If the binary is precessing, the GW amplitude and phase have a complicated dependency on the orientation of the binary, which provides additional information.

Our ability to characterize GW150914 as the signature of a binary system of compact objects, as we have outlined above, is dependent on the finite signal-to-noise ratio (SNR) of the signal and the specific properties of the underlying source. These properties described in detail below, and the inferred parameters for GW150914 are summarized in Table I and Figs. 1–6.

## II. METHOD

Full information about the properties of the source is provided by the probability density function (PDF) $p(\vec{\vartheta}|\vec{d})$ of the unknown parameters $\vec{\vartheta}$, given the two data streams from the instruments $\vec{d}$.

The posterior PDF is computed through a straightforward application of Bayes' theorem [40,41]. It is proportional to the product of the likelihood of the data given the parameters $\mathcal{L}(\vec{d}|\vec{\vartheta})$, and the prior PDF on the parameters $p(\vec{\vartheta})$ before we consider the data. From the (marginalized) posterior PDF, shown in Figs. 1–4 for selected parameters, we then construct credible intervals for the parameters, reported in Table I.

In addition, we can compute the evidence $\mathcal{Z}$ for the model under consideration. The evidence (also known as marginal likelihood) is the average of the likelihood under the prior on the unknown parameters for a specific model choice.

The computation of marginalized PDFs and the model evidence require the evaluation of multidimensional integrals. This is addressed by using a suite of Bayesian parameter-estimation and model-selection algorithms tailored to this problem [42]. We verify the results by using two *independent* stochastic sampling engines based on Markov-chain Monte Carlo [43,44] and nested sampling [5,45] techniques. (The marginalized PDFs and model evidence are computed using the `LALInference` package of the LIGO Algorithm Library (LAL) software suite [46]).

At the detector output we record the data $d_k(t) = n_k(t) + h_k^M(t;\vec{\vartheta})$, where $n_k$ is the noise, and $h_k^M$ is the measured strain, which differs from the physical strain $h_k$ from Eq. (1) as a result of the detectors' calibration [47]. In the frequency domain, we model the effect of calibration uncertainty by considering

$$\tilde{h}_k^M(f;\vec{\vartheta}) = \tilde{h}_k(f;\vec{\vartheta})[1+\delta A_k(f;\vec{\vartheta})] \times \exp[i\delta\phi_k(f;\vec{\vartheta})], \tag{4}$$

where $\tilde{h}_k^M$ and $\tilde{h}_k$ are the Fourier representation of the time-domain functions $h_k^M$ and $h_k$, respectively. $\delta A_k(f;\vec{\vartheta})$ and $\delta\phi_k(f;\vec{\vartheta})$ are the frequency-dependent amplitude and phase calibration-error functions, respectively. These calibration-error functions are modeled using a cubic spline polynomial, with five nodes per spline model placed uniformly in $\ln f$ [48].

We have analyzed the data at the time of this event using a *coherent* analysis. Under the assumption of stationary, Gaussian noise uncorrelated in each detector [49], the likelihood function for the LIGO network is [20,42]

$$\mathcal{L}(\vec{d}|\vec{\vartheta}) \propto \exp\left[-\frac{1}{2}\sum_{k=1,2}\langle h_k^M(\vec{\vartheta}) - d_k | h_k^M(\vec{\vartheta}) - d_k\rangle\right], \tag{5}$$

where $\langle\cdot|\cdot\rangle$ is the noise-weighted inner product [20]. We model the noise as a stationary Gaussian process of zero mean and known variance, which is estimated from the power spectrum computed using up to 1024 s of data adjacent to, but not containing, the GW signal [42].

The source properties are encoded into the two polarizations $h_+$ and $h_\times$ that enter the analysis through Eqs. (1) and (4). Here we focus on the case in which they originate from a compact binary coalescence; we use model waveforms (described below) that are based on solving Einstein's equations for the inspiral and merger of two BHs.

### A. BBH waveform models

For the modeled analysis of binary coalescences, an accurate waveform prediction for the gravitational radiation $h_{+,\times}$ is essential. As a consequence of the complexity of solving the two body problem in general relativity, several techniques have to be combined to describe all stages of the binary coalescence. While the early inspiral is well described by the analytical PN expansion [27], which relies on small velocities and weak gravitational fields, the strong-field merger stage can only be solved in full generality by large-scale NR simulations [32–34]. Since these pioneering works, numerous improvements have enabled numerical simulations of BBHs with sufficient accuracy for the applications considered here and for the region of parameter space of relevance to GW150914 (see, e.g., Refs. [50–52]). Tremendous progress has also been made in the past decade to combine analytical and numerical approaches, and now several accurate waveform models are available, and they are able to describe the entire coalescence for a large variety of possible configurations [51,53–59]. Extending and improving such models is an active area of research, and none of the current models can capture all possible physical effects (eccentricity, higher order gravitational modes in the presence of spins, etc.) for all conceivable binary systems. We discuss the current state of the art below.

In the Introduction, we outlined how the binary parameters affect the observable GW signal, and now we discuss the BH spins in greater detail. There are two main effects that the BH spins $\boldsymbol{S}_1$ and $\boldsymbol{S}_2$ have on the phase and amplitude evolution of the GW signal. The spin projections along the direction of the orbital angular momentum affect the inspiral rate of the binary. In particular, spin components aligned (antialigned) with $\boldsymbol{L}$ increase (decrease) the number of orbits from any given separation to merger with respect to the nonspinning case [27,60]. Given the limited SNR of the observed signal, it is difficult to untangle the full degrees of freedom of the individual BHs' spins, see, e.g., Refs. [61,62]. However, some spin information is encoded in a dominant spin effect. Several possible one-dimensional parametrizations of this effect can be found in the literature [21,63,64]; here, we use a simple mass-weighted linear combination of the spins [63,65–67]

$$\chi_{\rm eff} = \frac{c}{GM}\left(\frac{\boldsymbol{S}_1}{m_1} + \frac{\boldsymbol{S}_2}{m_2}\right)\cdot\frac{\boldsymbol{L}}{|\mathbf{L}|}, \tag{6}$$

which takes values between $-1$ (both BHs have maximal spins antialigned with respect to the orbital angular momentum) and $+1$ (maximal aligned spins).

Having described the effect of the two spin components aligned with the orbital angular momentum, four in-plane spin components remain. These lead to precession of the spins and the orbital plane, which in turn introduces modulations in the strain amplitude and phase as measured at the detectors. At leading order in the PN expansion, the equations that describe precession in a BBH are [30]

$$\dot{\boldsymbol{L}} = \frac{G}{c^2 r^3}(B_1\boldsymbol{S}_{1\perp} + B_2\boldsymbol{S}_{2\perp})\times\boldsymbol{L}, \tag{7}$$

$$\dot{\boldsymbol{S}}_i = \frac{G}{c^2 r^3}B_i\boldsymbol{L}\times\boldsymbol{S}_i, \tag{8}$$

where $\boldsymbol{S}_{i\perp}$ is the component of the spin perpendicular to $\boldsymbol{L}$; overdots denote time derivatives; $r$ is the orbital separation;

$B_1 = 2 + 3q/2$ and $B_2 = 2 + 3/(2q)$, and $i = \{1, 2\}$. It follows from Eqs. (7) and (8) that $\boldsymbol{L}$ and $\boldsymbol{S}_i$ precess around the almost-constant direction of the total angular momentum $\boldsymbol{J}$. For a nearly equal-mass binary (as we find is the case for GW150914), the precession angular frequency can be approximated by $\Omega_p \approx 7GJ/(c^2 r^3)$, and the total angular momentum is dominated during the inspiral by the orbital contribution, $J \approx L$. Additional, higher order spin-spin interactions can also contribute significantly to precession effects for some comparable-mass binaries [68,69].

The in-plane spin components rotate within the orbital plane at different velocities. Because of nutation of the orbital plane, the magnitude of the in-plane spin components oscillates around a mean value, but those oscillations are typically small. To first approximation, one can quantify the level of precession in a binary by averaging over the relative in-plane spin orientation. This is achieved by the following effective precession spin parameter [70]:

$$\chi_p = \frac{c}{B_1 G m_1^2} \max(B_1 S_{1\perp}, B_2 S_{2\perp}) > 0, \qquad (9)$$

where $\chi_p = 0$ corresponds to an aligned-spin (nonprecessing) system, and $\chi_p = 1$ to a binary with the maximum level of precession. Although the definition of $\chi_p$ is motivated by configurations that undergo many precession cycles during inspiral, we find that it is also a good approximate indicator of the in-plane spin contribution for the late inspiral and merger of GW150914.

For the analysis of GW150914, we consider waveform models constructed within two frameworks capable of accurately describing the GW signal in the parameter space of interest. The effective-one-body (EOB) formalism [71–75] combines perturbative results from the weak-field PN approximation with strong-field effects from the test-particle limit. The PN results are resummed to provide the EOB Hamiltonian, radiation-reaction force, and GW polarizations. The Hamiltonian is then improved by calibrating (unknown) higher order PN terms to NR simulations. Henceforth, we use "EOBNR" to indicate waveforms within this formalism. Waveform models that include spin effects have been developed, for both double nonprecessing spins [54,59] (11 independent parameters) and double precessing spins [55] (15 independent parameters). Here, we report results using the nonprecessing model [54] tuned to NR simulations [76]. This model is formulated as a set of differential equations that are computationally too expensive to solve for the millions of likelihood evaluations required for the analysis. Therefore, a frequency-domain reduced-order model [77,78] was implemented that faithfully represents the original model with an accuracy that is better than the statistical uncertainty caused by the instruments' noise. (In LAL, as well as in technical publications, the aligned, precessing and reduced-order EOBNR models are called `SEOBNRv2`, `SEOBNRv3` and `SEOBNRv2_ROM_DoubleSpin`, respectively). Bayesian analyses that use the double precessing spin model [55] are more time consuming and are not yet finalized. The results will be fully reported in a future publication.

An alternative inspiral-merger-ringdown phenomenological formalism [79–81] is based on extending frequency-domain PN expressions and hybridizing PN and EOB with NR waveforms. Henceforth, we use "IMRPhenom" to indicate waveforms within this formalism. Several waveform models that include aligned-spin effects have been constructed within this approach [58,66,67], and here we employ the most recent model based on a fitting of untuned EOB waveforms [54] and NR hybrids [57,58] of nonprecessing systems. To include leading-order precession effects, aligned-spin waveforms are rotated into precessing waveforms [82]. Although this model captures some two-spin effects, it was principally designed to accurately model the waveforms with respect to an effective spin parameter similar to $\chi_{\rm eff}$ above. The dominant precession effects are introduced through a lower-dimensional effective spin description [70,83], motivated by the same physical arguments as the definition of $\chi_p$. This provides us with an effective-precessing-spin model [83] with 13 independent parameters. (In LAL, as well as in some technical publications, the model used is called `IMRPhenomPv2`).

All models we use are restricted to circular inspirals. They also include only the dominant spherical harmonic modes in the nonprecessing limit.

### B. Choice of priors

We analyze coherently 8 s of data with a uniform prior on $t_c$ of width of $\pm 0.1$ s, centered on the time reported by the online analysis [1,84], and a uniform prior in $[0, 2\pi]$ for $\phi_c$. We consider the frequency region between 20 Hz, below which the sensitivity of the instruments significantly degrades (see panel (b) of Fig. 3 in Ref. [1]), and 1024 Hz, a safe value for the highest frequency contribution to radiation from binaries in the mass range considered here.

Given the lack of any additional astrophysical constraints on the source at hand, our prior choices on the parameters are uninformative. We assume sources uniformly distributed in volume and isotropically oriented. We use uniform priors in $m_{1,2} \in [10, 80] M_\odot$, with the constraint that $m_2 \leq m_1$. We use a uniform prior in the spin magnitudes $a_{1,2} \in [0, 1]$. For angles subject to change due to precession effects we give values at a reference GW frequency $f_{\rm ref} = 20$ Hz. We use isotropic priors on the spin orientation for the precessing model. For the nonprecessing model, the prior on the spin magnitudes may be interpreted as the dimensionless spin projection onto $\boldsymbol{L}$ having a uniform distribution $[-1, 1]$. This range includes binaries where the two spins are strongly antialigned relative to one another. Many such antialigned-spin comparable-mass systems are unstable to large-angle precession well before entering our sensitive band [85,86] and could not have formed from an asymptotically spin antialigned binary. We could exclude those systems if we believe the binary is not precessing. However, we do not make this assumption here and instead accept that the models can only extract limited spin information about a more general, precessing binary.

We also need to specify the prior ranges for the amplitude and phase error functions $\delta A_k(f;\vec{\vartheta})$ and $\delta\phi_k(f;\vec{\vartheta})$, see Eq. (5). The calibration during the time of observation of GW150914 is characterized by a 1-$\sigma$ statistical uncertainty of no more than 10% in amplitude and 10° in phase [1,47]. We use zero-mean Gaussian priors on the values of the spline at each node with widths corresponding to the uncertainties quoted above [48]. Calibration uncertainties therefore add 10 parameters per instrument to the model used in the analysis. For validation purposes we also considered an independent method that assumes frequency-independent calibration errors [87], and obtained consistent results.

## III. RESULTS

The results of the analysis using binary coalescence waveforms are posterior PDFs for the parameters describing the GW signal and the model evidence. A summary is provided in Table I. For the model evidence, we quote (the logarithm of) the Bayes factor $\mathcal{B}_{s/n} = \mathcal{Z}/\mathcal{Z}_n$, which is the evidence for a coherent signal hypothesis divided by that for (Gaussian) noise [5]. At the leading order, the Bayes factor and the optimal SNR $\rho = [\sum_k \langle h_k^M | h_k^M \rangle]^{1/2}$ are related by $\ln \mathcal{B}_{s/n} \approx \rho^2/2$ [88].

Before discussing parameter estimates in detail, we consider how the inference is affected by the choice of the compact-binary waveform model. From Table I, we see that the posterior estimates for each parameter are broadly consistent across the two models, despite the fact that they are based on different analytical approaches and that they include different aspects of BBH spin dynamics. The models' logarithms of the Bayes factors, $288.7 \pm 0.2$ and $290.3 \pm 0.1$, are also comparable for both models: the data do not allow us to conclusively prefer one model over the other [89]. Therefore, we use both for the Overall column in Table I. We combine the posterior samples of both distributions with equal weight, in effect marginalizing over our choice of waveform model. These averaged results give our best estimate for the parameters describing GW150914.

In Table I, we also indicate how sensitive our results are to our choice of waveform. For each parameter, we give systematic errors on the boundaries of the 90% credible intervals due to the uncertainty in the waveform models considered in the analysis; the quoted values are the 90% range of a normal distribution estimated from the variance of results from the different models. (If $X$ were an edge of a credible interval, we quote systematic uncertainty $\pm 1.64\sigma_{\rm sys}$ using the estimate $\sigma_{\rm sys}^2 = [(X_{\rm EOBNR} - X_{\rm Overall})^2 + (X_{\rm IMRPhenom} - X_{\rm Overall})^2]/2$. For parameters with bounded ranges, like the spins, the normal distributions should be truncated. However, for transparency, we still quote the 90% range of the uncut distributions. These numbers provide estimates of the order of magnitude of the potential systematic error). Assuming a normally distributed error is the least constraining choice [90] and gives a conservative estimate. The uncertainty from waveform modeling is less significant than the statistical uncertainty; therefore, we are confident that the results are robust against this potential systematic error. We consider this point in detail later in the Letter.

The analysis presented here yields an optimal coherent SNR of $\rho = 25.1^{+1.7}_{-1.7}$. This value is higher than the one reported by the search [1,3] because it is obtained using a finer sampling of (a larger) parameter space.

GW150914's source corresponds to a stellar-mass BBH with individual source-frame masses $m_1^{\rm source} = 36^{+5}_{-4} M_\odot$ and $m_2^{\rm source} = 29^{+4}_{-4} M_\odot$, as shown in Table I and Fig. 1. The two BHs are nearly equal mass. We bound the mass ratio to the range $0.66 \leq q \leq 1$ with 90% probability. For comparison, the highest observed neutron star mass is $2.01 \pm 0.04 M_\odot$ [91], and the conservative upper-limit for the mass of a stable neutron star is $3M_\odot$ [92,93]. The masses inferred from GW150914 are an order of magnitude larger than these values, which implies that these two compact objects of GW150914 are BHs, unless exotic alternatives, e.g., boson stars [94], do exist. If the compact objects were not BHs, this would leave an imprint on the waveform, e.g., Ref. [95]; however, in Ref. [96] we verify that the observed signal is consistent with that predicted assuming BHs in general relativity. These results establish the presence of stellar-mass BBHs in the Universe. It also proves that BBHs formed in nature can merge within a Hubble time [97].

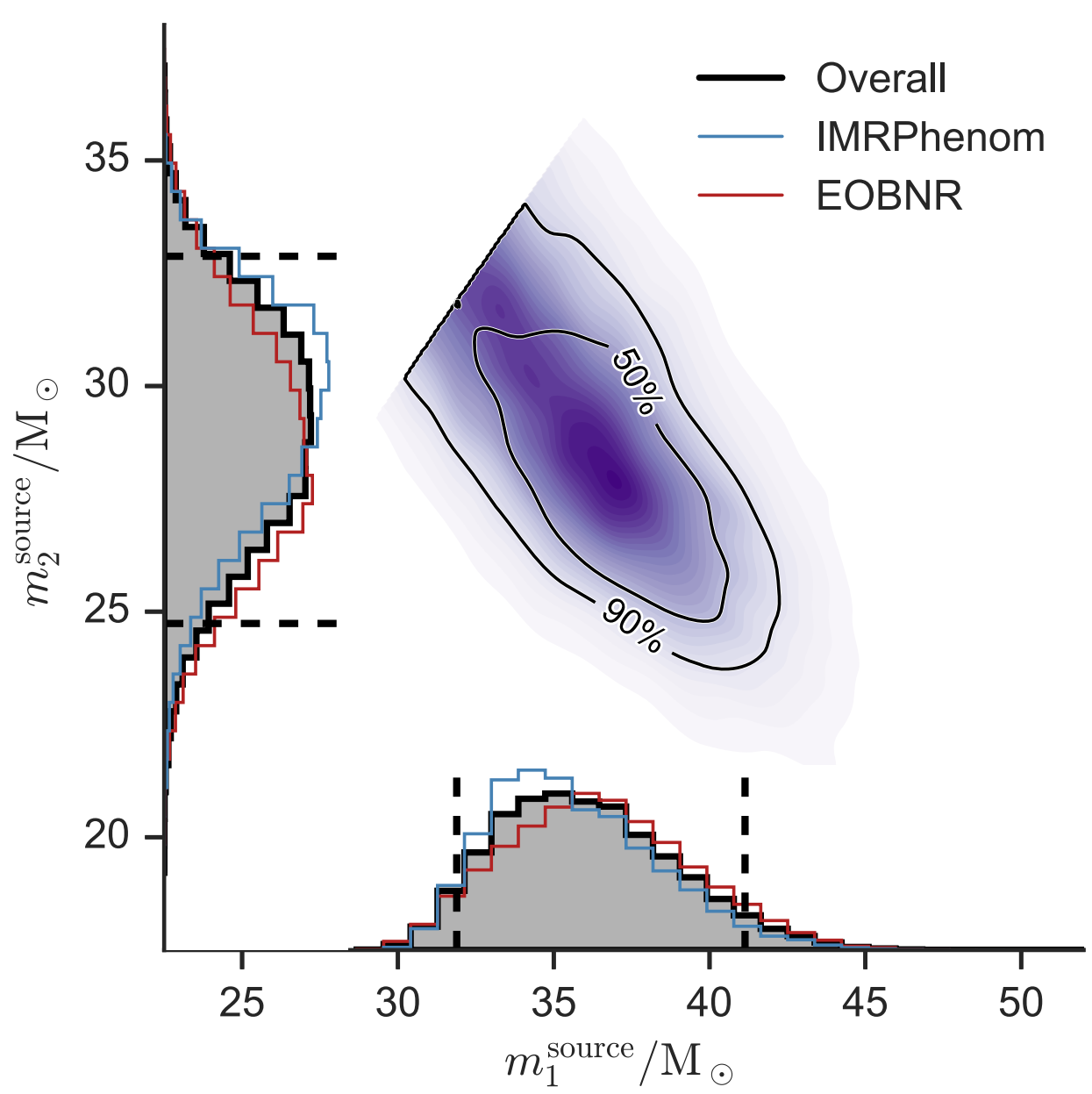

FIG. 1. Posterior PDFs for the source-frame component masses $m_1^{\rm source}$ and $m_2^{\rm source}$. We use the convention that $m_2^{\rm source} \leq m_1^{\rm source}$, which produces the sharp cut in the two-dimensional distribution. In the one-dimensional marginalized distributions we show the Overall (solid black), IMRPhenom (blue), and EOBNR (red) PDFs; the dashed vertical lines mark the 90% credible interval for the Overall PDF. The two-dimensional plot shows the contours of the 50% and 90% credible regions plotted over a color-coded PDF.

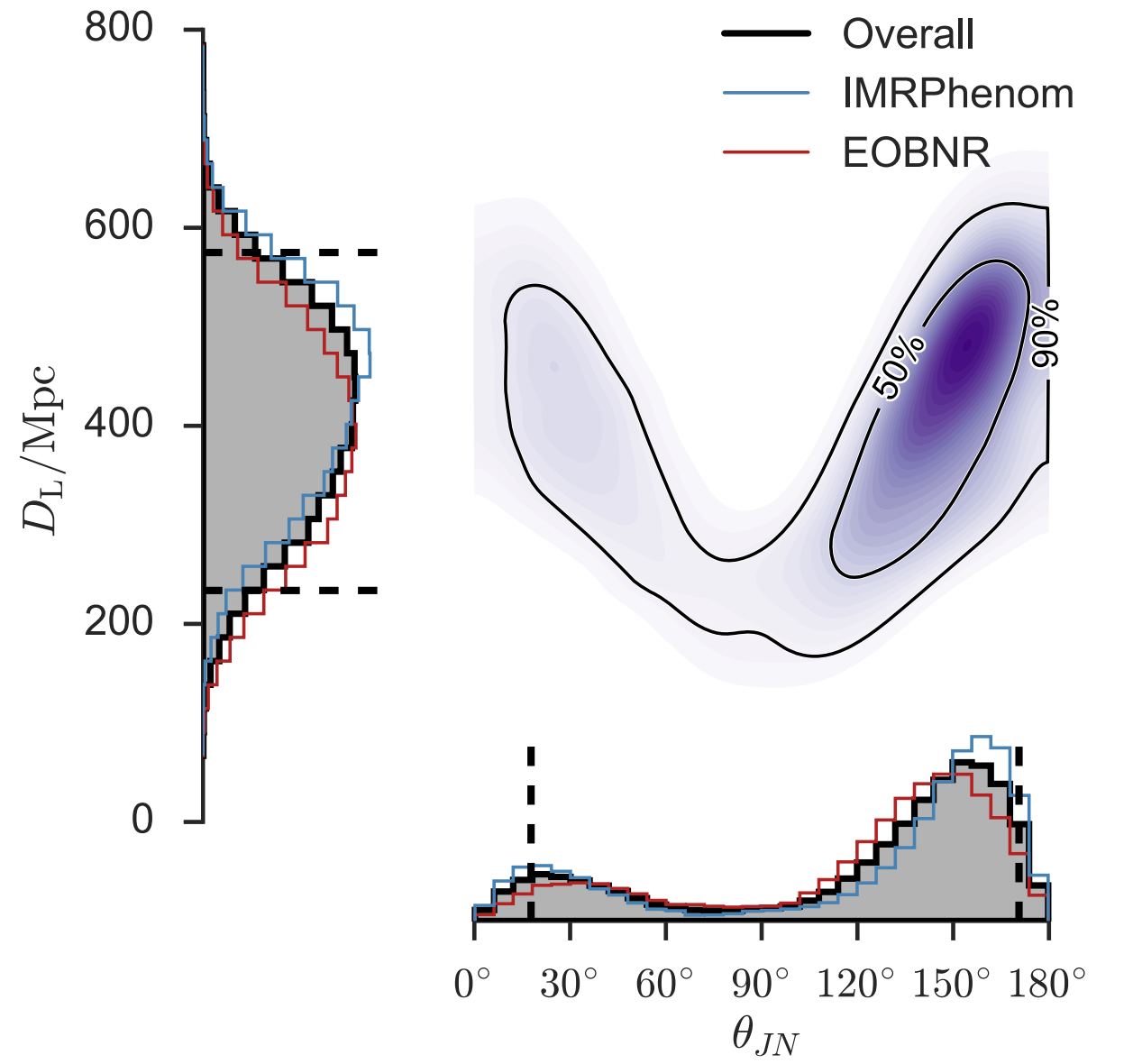

FIG. 2. Posterior PDFs for the source luminosity distance $D_L$ and the binary inclination $\theta_{JN}$. In the one-dimensional marginalized distributions we show the Overall (solid black), IMRPhenom (blue), and EOBNR (red) PDFs; the dashed vertical lines mark the 90% credible interval for the Overall PDF. The two-dimensional plot shows the contours of the 50% and 90% credible regions plotted over a color-coded PDF.

To convert the masses measured in the detector frame to physical source-frame masses, we require the redshift of the source. As discussed in the Introduction, GW observations are directly sensitive to the luminosity distance to a source, but not the redshift [98]. We find that GW150914 is at $D_L = 410^{+160}_{-180}$ Mpc. Assuming a flat ΛCDM cosmology with Hubble parameter $H_0 = 67.9\ \mathrm{km\,s^{-1}\,Mpc^{-1}}$ and matter density parameter $\Omega_m = 0.306$ [6], the inferred luminosity distance corresponds to a redshift of $z = 0.09^{+0.03}_{-0.04}$.

The luminosity distance is strongly correlated to the inclination of the orbital plane with respect to the line of sight [4,20,99]. For precessing systems, the orientation of the orbital plane is time dependent. We therefore describe the source inclination by $\theta_{JN}$, the angle between the total angular momentum (which typically is approximately constant throughout the inspiral) and the line of sight [30,100], and we quote its value at a reference GW frequency $f_{\rm ref} = 20$ Hz. The posterior PDF shows that an orientation of the total orbital angular momentum of the BBH strongly misaligned to the line of sight is disfavored; the probability that $45° < \theta_{JN} < 135°$ is 0.35.

The masses and spins of the BHs in a (circular) binary are the only parameters needed to determine the final mass and spin of the BH that is produced at the end of the merger. Appropriate relations are embedded intrinsically in the waveform models used in the analysis, but they do not give direct access to the parameters of the remnant BH. However, applying the fitting formula calibrated to nonprecessing NR simulations provided in Ref. [101] to the posterior for the component masses and spins [102], we infer the mass and spin of the remnant BH to be $M_f^{\rm source} = 62^{+4}_{-4} M_\odot$, and $a_f = 0.67^{+0.05}_{-0.07}$, as shown in Fig. 3 and Table I. These results are fully consistent with those obtained using an independent nonprecessing fit [57]. The systematic uncertainties of the fit are much smaller than the statistical uncertainties. The value of the final spin is a consequence of conservation of angular momentum in which the total angular momentum of the system (which for a nearly equal mass binary, such as GW150914's source, is dominated by the orbital angular momentum) is converted partially into the spin of the remnant black hole and partially radiated away in GWs during the merger. Therefore, the final spin is more precisely determined than either of the spins of the binary's BHs.

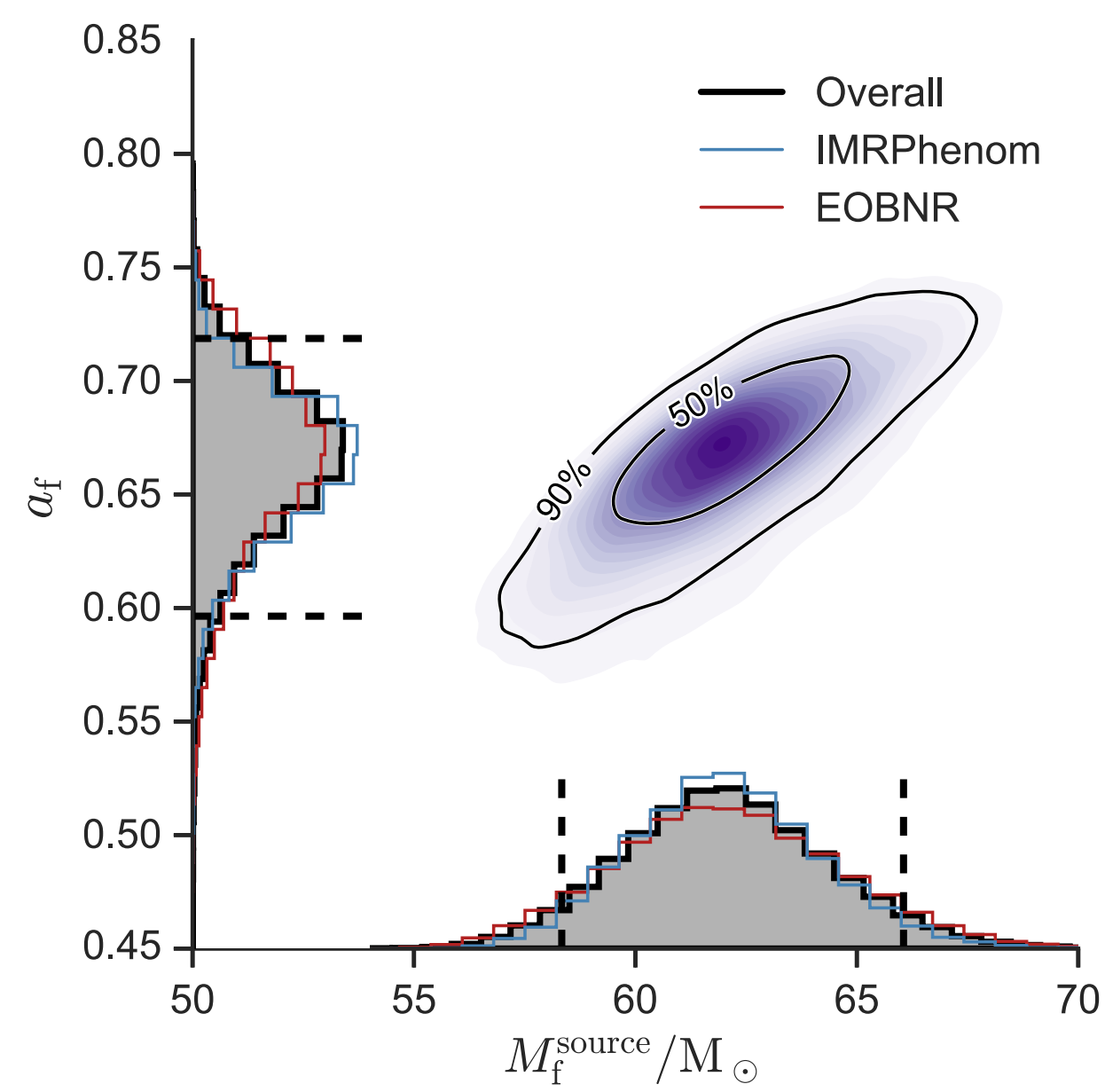

FIG. 3. PDFs for the source-frame mass and spin of the remnant BH produced by the coalescence of the binary. In the one-dimensional marginalized distributions we show the Overall (solid black), IMRPhenom (blue), and EOBNR (red) PDFs; the dashed vertical lines mark the 90% credible interval for the Overall PDF. The two-dimensional plot shows the contours of the 50% and 90% credible regions plotted over a color-coded PDF.

The calculation of the final mass also provides an estimate of the total energy radiated in GWs as $E_{\rm rad} = M^{\rm source} - M_f^{\rm source}$. GW150914 emitted a total of $E_{\rm rad} = 3.0^{+0.5}_{-0.4} M_\odot c^2 = 5.3^{+0.9}_{-0.8} \times 10^{47}$J in GWs, the majority of which was at frequencies in LIGO's sensitive band. These values are fully consistent with those given in the literature for NR simulations of similar binaries [103,104]. The energetics of a BBH merger can be estimated at the order of magnitude level using simple Newtonian arguments. The total energy of a binary system at separation $r$ is given by $E \approx Mc^2 - Gm_1m_2/(2r)$. For an equal-mass system, and assuming the inspiral phase to end at about $r \approx 5GM/c^2$, then around 2%–3% of the initial total energy of the system is emitted as GWs. Only a fully general relativistic treatment of the system can accurately describe the physical process during the final strong-field phase of the coalescence. This indicates that a comparable amount of energy is emitted during the merger portion of GW150914, leading to ≈5% of the total energy emitted.

We further infer the peak GW luminosity achieved during the merger phase by applying to the posteriors a separate fit to nonprecessing NR simulations [105]. The source reached a maximum instantaneous GW luminosity of $3.5^{+0.5}_{-0.4} \times 10^{56}$ erg s$^{-1}$ $= 200^{+30}_{-20} M_\odot c^2$ s$^{-1}$. Here, the uncertainties include an estimate for the systematic error of the fit as obtained by comparison with a separate set of precessing NR simulations, in addition to the dominant statistical contribution. An order-of-magnitude estimate of the luminosity corroborates this result. For the dominant mode, the flux can be estimated by $\approx c^3|\dot{h}|^2/(16\pi G) \sim 10^5$ erg s$^{-1}$ m$^{-2}$, where $\dot{h}$ is the time derivative of the strain (cf. Ref. [106], [section 18.6]), and we use a GW amplitude of $|h| \approx 10^{-21}$ at a frequency of 250 Hz [1]. Using the inferred distance leads to an estimated luminosity of $\sim 10^{56}$ erg s$^{-1}$. For comparison, the ultraluminous GRB 110918A reached a peak isotropic-equivalent luminosity of $(4.7 \pm 0.2) \times 10^{54}$ erg s$^{-1}$ [107].

GW ground-based instruments are all-sky monitors with no intrinsic spatial resolution capability for transient signals. A network of instruments is needed to reconstruct the location of a GW in the sky, via time of arrival, and amplitude and phase consistency across the network [108]. The observed time delay of GW150914 between the Livingston and Hanford observatories was $6.9^{+0.5}_{-0.4}$ ms. With only the two LIGO instruments in observational mode, GW150914's source location can only be reconstructed to approximately an annulus set to first approximation by this time delay [109–111]. Figure 4 shows the sky map for GW150914: it corresponds to a projected two-dimensional credible region of 150 deg$^2$ (50% probability) and 610 deg$^2$ (90% probability). The associated three-dimensional comoving volume probability region is $\sim 10^{-2}$Gpc$^3$; for comparison the comoving density of Milky Way-equivalent galaxies is $\sim 10^7$Gpc$^{-3}$. This area of the sky was targeted by follow-up observations covering radio, optical, near infrared, x-ray, and gamma-ray wavelengths that are discussed in Ref. [112]; searches for coincident neutrinos are discussed in Ref. [113].

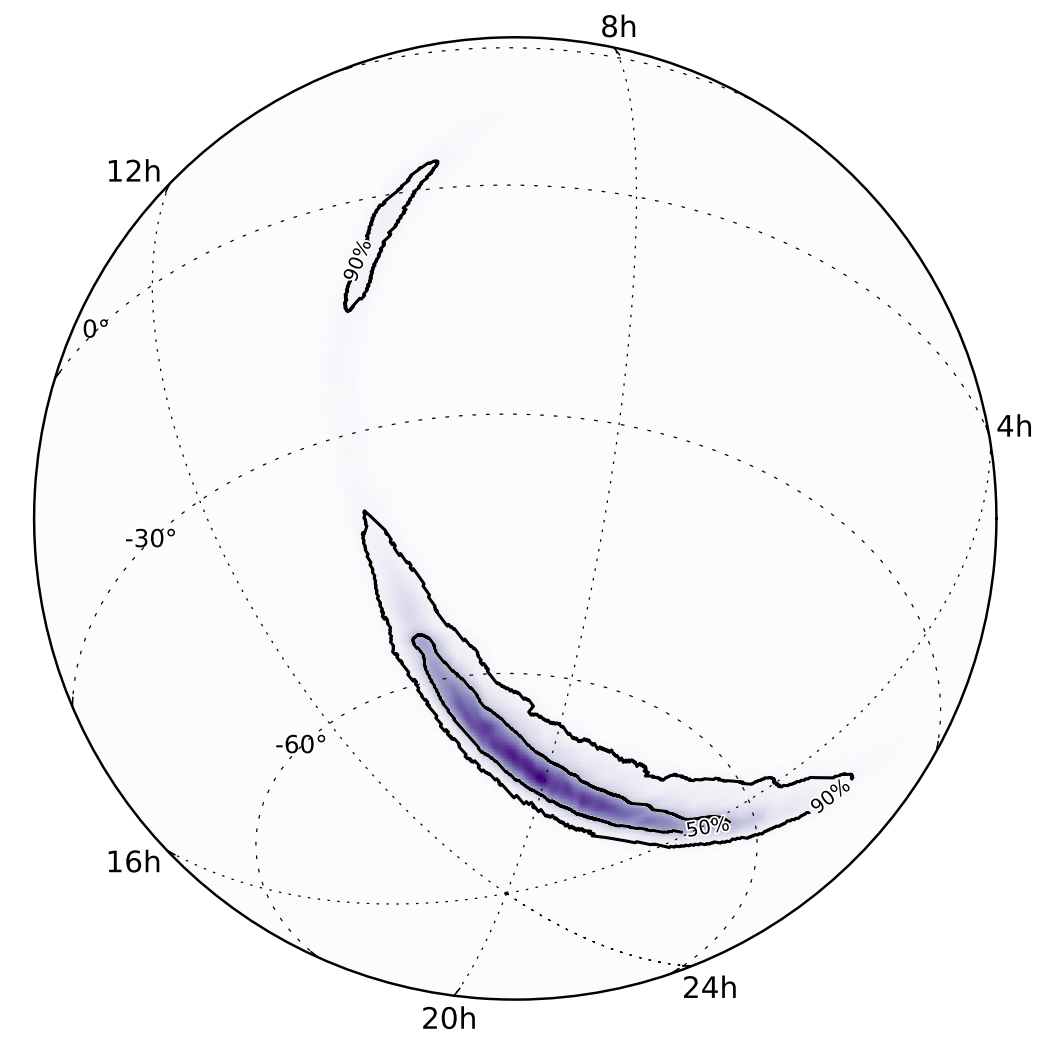

FIG. 4. An orthographic projection of the PDF for the sky location of GW150914 given in terms of right ascension $\alpha$ (measured in hours and labeled around the edge of the figure) and declination $\delta$ (measured in degrees and labeled inside the figure). The contours of the 50% and 90% credible regions are plotted over a color-coded PDF. The sky localization forms part of an annulus, set by the time delay of $6.9^{+0.5}_{-0.4}$ ms between the Livingston and Hanford detectors.

Spins are a fundamental property of BHs. Additionally, their magnitude and orientation with respect to the orbital angular momentum carry an imprint of the evolutionary history of a binary that could help in identifying the formation channel, such as distinguishing binaries formed in the field from those produced through captures in dense stellar environments [97]. The observation of GW150914 allows us for the first time to put direct constraints on BH spins. The EOBNR and IMRPhenom models yield consistent values for the magnitude of the individual spins, see Table I. The spin of the primary BH is constrained to $a_1 < 0.7$ (at 90% probability), and strongly disfavors the primary BH being maximally spinning. The bound on the secondary BH's spin is $a_2 < 0.9$ (at 90% probability), which is consistent with the bound derived from the prior.

Results for precessing spins are derived using the IMRPhenom model. Spins enter the model through the two effective spin parameters $\chi_{\rm eff}$ and $\chi_p$. The left panel of Fig. 5 shows that despite the short duration of the signal in band we meaningfully constrain $\chi_{\rm eff} = -0.07^{+0.16}_{-0.17}$, see Table I. The inspiral rate of GW150914 is therefore only weakly affected by the spins. We cannot, however, extract additional information on the other spin components associated with precession effects. The data are uninformative: the posterior PDF on $\chi_p$ (left panel of Fig. 5) is broadly consistent with the prior, and the distribution of spins (right panel of Fig. 5) matches our expectations once the information that $|\chi_{\rm eff}|$ is small has been included. Two elements may be responsible for this. *If* precession occurs, at most one modulation cycle would be present in the LIGO sensitivity window. *If* the source was viewed with $\boldsymbol{J}$ close to the line of sight (Fig. 2), the amplitude of possible modulations in the recorded strain is suppressed.

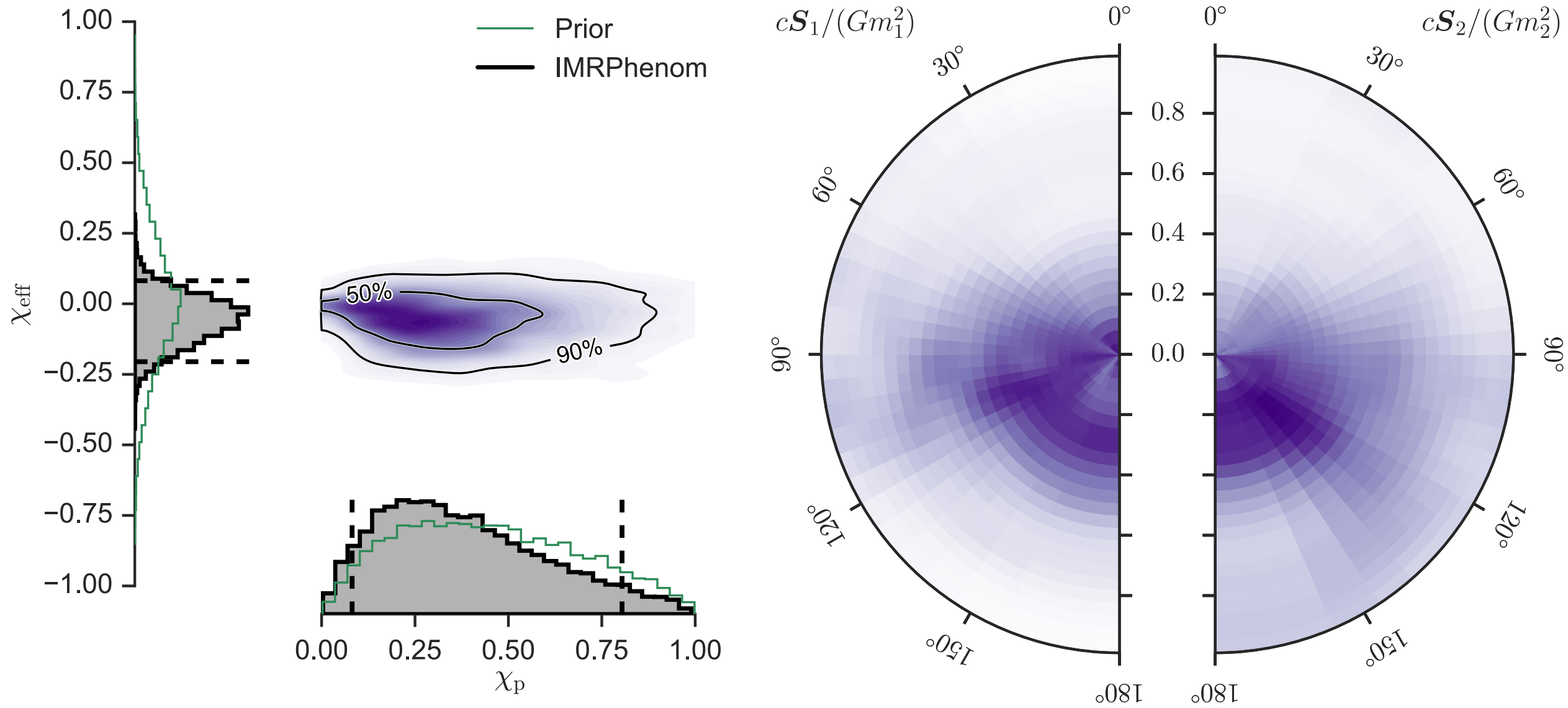

FIG. 5. Left: PDFs (solid black line) for the $\chi_p$ and $\chi_{\rm eff}$ spin parameters compared to their prior distribution (green line). The dashed vertical lines mark the 90% credible interval. The one-dimensional plots show probability contours of the prior (green) and marginalized PDF (black). The two-dimensional plot shows the contours of the 50% and 90% credible regions plotted over a color-coded PDF. Right: PDFs for the dimensionless component spins $c\boldsymbol{S}_1/(Gm_1^2)$ and $c\boldsymbol{S}_2/(Gm_2^2)$ relative to the normal to the orbital plane $\boldsymbol{L}$, marginalized over uncertainties in the azimuthal angles. The bins are constructed linearly in spin magnitude and the cosine of the tilt angles, $\cos\theta_{LS_i} = \boldsymbol{S}_i \cdot \boldsymbol{L}/(|\boldsymbol{S}_i||\boldsymbol{L}|)$, where $i = \{1, 2\}$, and therefore have equal prior probability.

The joint posterior PDFs of the magnitude and orientation of $\boldsymbol{S}_1$ and $\boldsymbol{S}_2$ are shown in the right panel of Fig. 5. The angle of the spins with respect to $\boldsymbol{L}$ (the tilt angle) is considered a tracer of BBH formation channels [97]. However, we can place only weak constraints on this parameter for GW150914: the probabilities that $\boldsymbol{S}_1$ and $\boldsymbol{S}_2$ are at an angle between 45° and 135° with respect to the normal to the orbital plane $\boldsymbol{L}$ are 0.77 and 0.75, respectively. For this specific geometrical configuration the spin magnitude estimates are $a_1 < 0.8$ and $a_2 < 0.8$ at 90% probability.

Some astrophysical formation scenarios favor spins nearly aligned with the orbital angular momentum, particularly for the massive progenitors that in these scenarios produce GW150914 [97,114,115]. To estimate the impact of this prior hypothesis on our interpretation, we used the fraction (2.5%) of the spin-aligned result (EOBNR) with $\boldsymbol{S}_{1,2} \cdot \boldsymbol{L} > 0$ to revise our expectations. If both spins must be positively and strictly co-aligned with $\boldsymbol{L}$, then we can constrain the two individual spins at 90% probability to be $a_1 < 0.2$ and $a_2 < 0.3$.

The loss of linear momentum through GWs produces a recoil of the merger BH with respect to the binary's original center of mass [116,117]. The recoil velocity depends on the spins (magnitude and orientation) of the BHs of the binary and could be large for spins that are appropriately misaligned with the orbital angular momentum [118–121]. Unfortunately, the weak constraints on the spins (magnitude and direction) of GW150914 prevent us from providing a meaningful limit on the kick velocity of the resulting BH.

### A. A minimal-assumption analysis

In addition to the analysis based on the assumption that the signal is generated by a binary system, we also consider a model which is not derived from a particular physical scenario and makes minimal assumptions about $h_{+,\times}$. In this case we compute directly the posterior $p(\vec{h}|\vec{d})$ by reconstructing $h_{+,\times}$ using a linear combination of elliptically polarized sine-Gaussian wavelets whose amplitudes are assumed to be consistent with a uniform source distribution [84,122], see Fig. 6. The number of wavelets in the linear combination is not fixed *a priori* but is optimized via Bayesian model selection. This analysis directly infers the PDF of the GW strain given the data $p(\vec{h}|\vec{d})$.

We can compare the minimal-assumption posterior for the strain at the two instruments with the results of the compact binary modeled analysis $p(\vec{h}(\vec{\vartheta})|\vec{d})$. The waveforms are shown in Fig. 6. There is remarkable agreement between the actual data and the reconstructed waveform under the two model assumptions. As expected, the uncertainty is greater for the minimal-assumption reconstruction due to greater flexibility in its waveform model. The agreement between the reconstructed waveforms using the two models can be quantified through the noise-weighted inner product that enters Eq. (5), and it is found to be $94^{+2}_{-3}\%$, consistent with expectations for the SNR at which GW150914 was observed.

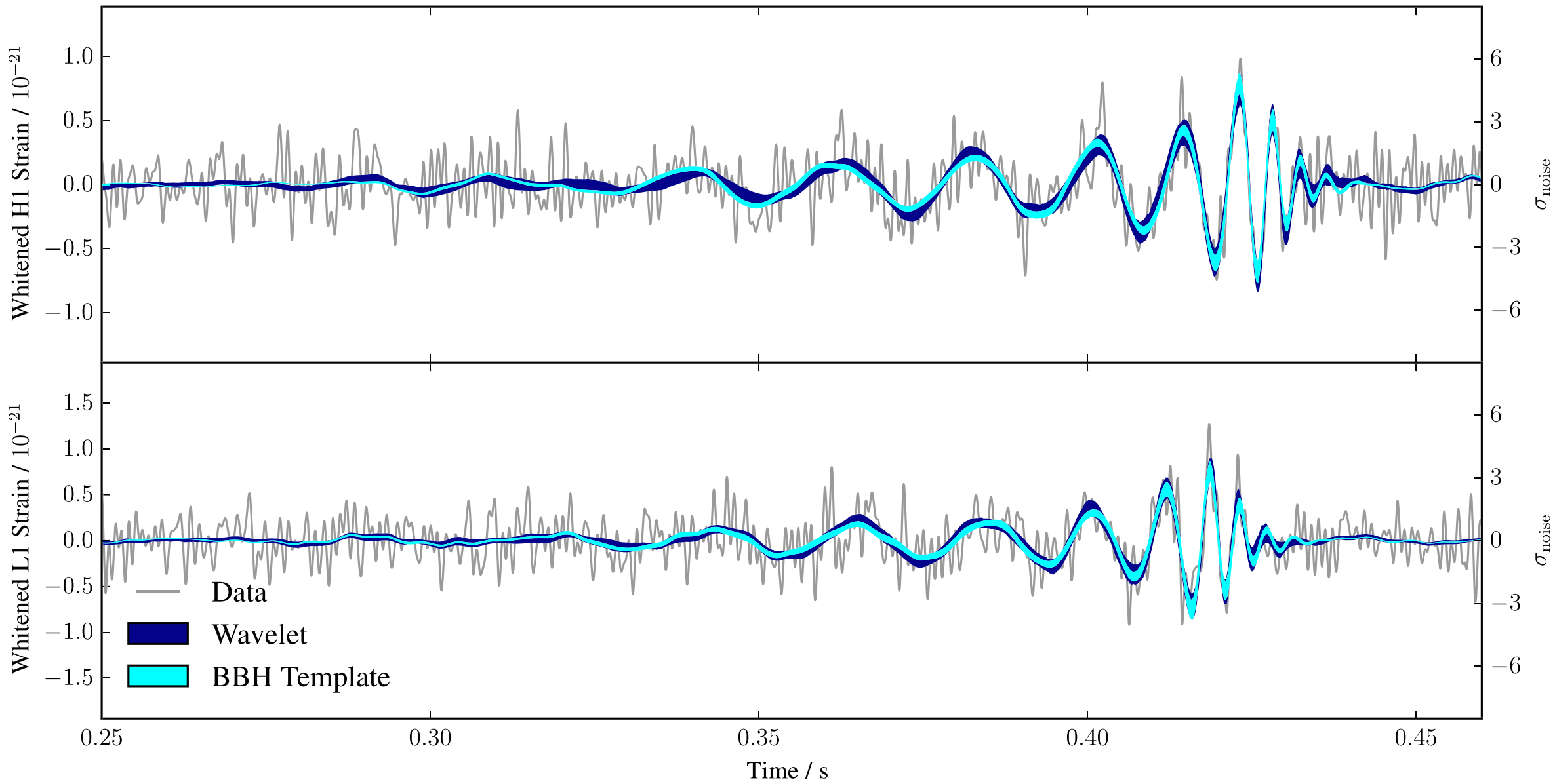

FIG. 6. Time-domain data (sampled at 2048 Hz) and reconstructed waveforms of GW150914, whitened by the noise power spectral density (in Fig. 1 of Ref. [1] the data are band passed and notched filtered), for the H1 (top) and L1 (bottom) detectors. Times are shown relative to September 14, 2015 at 09:50:45 UTC. The ordinate axes on the right are in units of noise standard deviations from zero—i.e., the peak alone is an $\sim$4-$\sigma$ excursion relative to the instrument noise at that time—and on the left are normalized in order to preserve the strain amplitude at 200 Hz. The waveforms are plotted as bands representing the uncertainty in the reconstruction. Shaded regions correspond to the 90% credible regions for the reconstructed waveforms. The broadest (dark blue) shaded region is obtained with the model that does not assume a particular waveform morphology, but instead uses a linear combination of sine-Gaussian wavelets. The lighter, narrower shaded region (cyan) is the result from the modeled analyses using IMRPhenom and EOBNR template waveforms. The thin grey lines are the data. The agreement between the reconstructed waveforms using the two models is found to be $94^{+2}_{-3}\%$.

## IV. DISCUSSION

We have presented measurements of the heaviest stellar-mass BHs known to date, and the first stellar-mass BBH. The system merges into a BH of $\approx 60 M_\odot$. So far, stellar-mass BHs of masses $\approx 10 M_\odot$ have been claimed using dynamical measurement of Galactic x-ray binaries [123]. Masses as high as 16–20$M_\odot$ and 21–35$M_\odot$ have been reported for IC10 X-1 [124,125] and NGC300 X-1 [126], respectively; however, these measurements may have been contaminated by stellar winds as discussed in Ref. [127] and references therein. Our results attest that BBHs do form and merge within a Hubble time. We have constrained the spin of the primary BH of the binary to be $a_1 < 0.7$ and we have inferred the spin of the remnant BH to be $a_f \approx 0.7$. Up to now, spin estimates of BH candidates have relied on modeling of accretion disks to interpret spectra of x-ray binaries [128]. In contrast, GW measurements rely only on the predictions of general relativity for vacuum spacetime. Astrophysical implications of our results on the rate of BBH mergers are discussed in Ref. [129] and implications for our understanding of the formation and evolution of BBHs are discussed in Ref. [97].

The statistical uncertainties with which we have characterized the source properties and parameters reflect the finite SNR of the observation of GW150914 and the error budget of the strain calibration process. The latter degrades primarily the estimate of the source location. If we *assume* that the strain was perfectly calibrated, i.e., $h^M = h$, see Eqs. (1) and (4), the 50% and 90% credible regions for sky location would become 48 deg$^2$ and 150 deg$^2$, compared to the actual results of 150 deg$^2$ and 610 deg$^2$, respectively. The physical parameters show only small changes with the marginalization over calibration uncertainty, for example, the final mass $M_f^{\text{source}}$ changes from $62^{+4}_{-4} M_\odot$ including calibration uncertainty to $62^{+4}_{-3} M_\odot$ assuming perfect calibration, and the final spin $a_f$ changes from $0.67^{+0.05}_{-0.07}$ to $0.67^{+0.04}_{-0.05}$. The effect of calibration uncertainty is to increase the overall parameter range at a given probability, but the medians of the PDFs remain largely unchanged. For GW150914, the dominant source of statistical uncertainty is the finite SNR. More accurate calibration techniques are currently being tested, and one can expect that in future

LIGO observations the impact of calibration errors on the inference of GW source parameters will be further reduced [47].

Our analysis is based on waveform models that do not include the effect of spins in their full generality. Still, the aligned-spin EOBNR and precessing IMRPhenom models produce consistent results, with an error budget dominated by statistical uncertainty and not systematic errors as evaluated from the physics encoded in these two models. It is, however, important to consider whether *additional* systematics produced by physics not captured by these models could have affected the results. To this extent, we collected existing numerical waveforms and generated targeted new simulations [130–135]. The simulations were generated with multiple independent codes and sample the posterior region for masses and spins inferred for GW150914. We have added these NR waveforms as mock signals to the data in the neighborhood of GW150914 and to synthetic data sets representative of LIGO's sensitivity and noise properties at the time of GW150914, with SNRs consistent with the one reported for the actual detection. We then carried out exactly the same analysis applied to the actual data.

For signals from nonprecessing binaries, we recover parameters that are consistent with those describing the mock source. The results obtained with the EOBNR and IMRPhenom model are consistent, which further confirms previous comparison studies of nonprecessing models with NR waveforms [54,58,136]. For signals that describe precessing binaries, but with orbital angular momentum orientation consistent with the most likely geometry inferred for GW150914, i.e., orbital angular momentum close to aligned or antialigned with the line of sight, we find again that the PDFs are consistent across the models and with the true values of the parameters used for the numerical simulation. For the same physical parameters, but a total angular momentum orientated to give the largest amount of signal modulations at the instrument output, i.e., $\boldsymbol{J}$ approximately perpendicular to the line of sight, the results using the EOBNR and IMRPhenom models do differ from each other. They yield biased and statistically inconsistent PDFs, depending on the specific NR configuration used as the mock signal. This is partly due to the fact that not all physical effects are captured by the models (as in the case of the nonprecessing EOBNR model) and partly due to systematic inaccuracies of the models. However, we stress that this is not the configuration we find for GW150914.

The outcome of these studies suggests that for GW150914, the results reported here are not appreciably affected by additional waveform systematics beyond those quantified in Table I. A detailed analysis will be presented in a forthcoming paper. In addition, we are currently carrying out an analysis using generalized, precessing EOBNR waveforms [55], which depend on the full 15 independent parameters of a coalescing binary in circular orbit. Preliminary investigations give results that are broadly consistent with those presented here based on the precessing IMRPhenom model, and full details will be reported in the future.

Throughout this work we have considered a model for the binary evolution in the LIGO sensitivity band that assumes a circular orbit. The posterior waveforms according to this model are consistent with minimal-assumption reconstructed waveforms, which make no assumption about eccentricity. Preliminary investigations suggest that eccentricities of $e \lesssim 0.1$ at 10 Hz would not introduce measurable deviations from a circular-orbit signal; however, even larger eccentricities may have negligible effects on the recovered source parameters. At this time, the lack of a model that consistently accounts for the presence of spins and eccentricity throughout the full coalescence prevents us from placing more stringent constraints. We plan to report improved limits in the future.

The analysis reported here is carried out under the assumption that general relativity is correct. The results presented in Fig. 6 show no evidence for deviations from the null hypothesis general-relativity model. However, GW150914 provides a new arena for tests of Einstein's theory in unexplored regimes; these are discussed in detail in Ref. [96].

## V. SUMMARY

We have reported the properties of GW150914 derived from a coherent analysis of data from the two LIGO detectors, based on accurate modeling of the coalescence signal as predicted by general relativity. We have shown that GW150914 originates from a BBH system with component masses $36^{+5}_{-4} M_\odot$ and $29^{+4}_{-4} M_\odot$ at redshift $0.09^{+0.03}_{-0.04}$ that merges to form a BH of mass $62^{+4}_{-4} M_\odot$ and spin $0.67^{+0.05}_{-0.07}$. The final BH is more massive than any other found in the stellar-mass range. We bound the spin magnitude of the binary's primary BH to <0.7 (at 90% probability), and measure the effective spin parameter along the direction of the orbital angular momentum to be $-0.07^{+0.16}_{-0.17}$, but we cannot place meaningful limits on the precession effects. Further implications stemming from our findings for tests of general relativity, the rate of BBH mergers, and the astrophysics of the BBHs are reported in Refs. [96,129], and [97], respectively. These results herald the beginning of GW astronomy and provide the first observational insights into the physics of BBHs.

## ACKNOWLEDGMENTS

The authors gratefully acknowledge the support of the United States National Science Foundation (NSF) for the construction and operation of the LIGO Laboratory and Advanced LIGO as well as the Science and Technology Facilities Council (STFC) of the United Kingdom, the Max-Planck-Society (MPS), and the State of Niedersachsen, Germany for support of the construction of Advanced LIGO and construction and operation of the GEO 600 detector. Additional support for Advanced LIGO was provided by the Australian Research Council. The authors gratefully acknowledge the Italian Istituto Nazionale di Fisica Nucleare (INFN), the French Centre National de la Recherche Scientifique (CNRS) and the Foundation for Fundamental Research on Matter supported by the Netherlands Organisation for Scientific Research, for the construction and operation of the Virgo detector and the creation and support of the EGO consortium. The authors also gratefully acknowledge research support from these agencies as well as by the Council of Scientific and Industrial Research of India, Department of Science and Technology, India, Science & Engineering Research Board (SERB), India, Ministry of Human Resource Development, India, the Spanish Ministerio de Economía y Competitividad, the Conselleria d'Economia i Competitivitat and Conselleria d'Educació, Cultura i Universitats of the Govern de les Illes Balears, the National Science Centre of Poland, the European Commission, the Royal Society, the Scottish Funding Council, the Scottish Universities Physics Alliance, the Hungarian Scientific Research Fund (OTKA), the Lyon Institute of Origins (LIO), the National Research Foundation of Korea, Industry Canada and the Province of Ontario through the Ministry of Economic Development and Innovation, the Natural Science and Engineering Research Council Canada, Canadian Institute for Advanced Research, the Brazilian Ministry of Science, Technology, and Innovation, Russian Foundation for Basic Research, the Leverhulme Trust, the Research Corporation, Ministry of Science and Technology (MOST), Taiwan and the Kavli Foundation. The authors gratefully acknowledge the support of the NSF, STFC, MPS, INFN, CNRS and the State of Niedersachsen, Germany for provision of computational resources.

B. P. Abbott,[1] R. Abbott,[1] T. D. Abbott,[2] M. R. Abernathy,[1] F. Acernese,[3,4] K. Ackley,[5] C. Adams,[6] T. Adams,[7] P. Addesso,[3] R. X. Adhikari,[1] V. B. Adya,[8] C. Affeldt,[8] M. Agathos,[9] K. Agatsuma,[9] N. Aggarwal,[10] O. D. Aguiar,[11] L. Aiello,[12,13] A. Ain,[14] P. Ajith,[15] B. Allen,[8,16,17] A. Allocca,[18,19] P. A. Altin,[20] S. B. Anderson,[1] W. G. Anderson,[16] K. Arai,[1] M. C. Araya,[1] C. C. Arceneaux,[21] J. S. Areeda,[22] N. Arnaud,[23] K. G. Arun,[24] S. Ascenzi,[25,13] G. Ashton,[26] M. Ast,[27] S. M. Aston,[6] P. Astone,[28] P. Aufmuth,[8] C. Aulbert,[8] S. Babak,[29] P. Bacon,[30] M. K. M. Bader,[9] P. T. Baker,[31] F. Baldaccini,[32,33] G. Ballardin,[34] S. W. Ballmer,[35] J. C. Barayoga,[1] S. E. Barclay,[36] B. C. Barish,[1] D. Barker,[37] F. Barone,[3,4] B. Barr,[36] L. Barsotti,[10] M. Barsuglia,[30] D. Barta,[38] J. Bartlett,[37] I. Bartos,[39] R. Bassiri,[40] A. Basti,[18,19] J. C. Batch,[37] C. Baune,[8] V. Bavigadda,[34] M. Bazzan,[41,42] B. Behnke,[29] M. Bejger,[43] A. S. Bell,[36] C. J. Bell,[36] B. K. Berger,[1] J. Bergman,[37] G. Bergmann,[8] C. P. L. Berry,[44] D. Bersanetti,[45,46] A. Bertolini,[9] J. Betzwieser,[6] S. Bhagwat,[35] R. Bhandare,[47] I. A. Bilenko,[48] G. Billingsley,[1] J. Birch,[6] R. Birney,[49] O. Birnholtz,[8] S. Biscans,[10] A. Bisht,[8,17] M. Bitossi,[34] C. Biwer,[35] M. A. Bizouard,[23] J. K. Blackburn,[1] C. D. Blair,[50] D. G. Blair,[50] R. M. Blair,[37] S. Bloemen,[51] O. Bock,[8] T. P. Bodiya,[10] M. Boer,[52] G. Bogaert,[52] C. Bogan,[8] A. Bohe,[29] P. Bojtos,[53] C. Bond,[44] F. Bondu,[54] R. Bonnand,[7] B. A. Boom,[9] R. Bork,[1] V. Boschi,[18,19] S. Bose,[55,14] Y. Bouffanais,[30] A. Bozzi,[34] C. Bradaschia,[19] P. R. Brady,[16] V. B. Braginsky,[48] M. Branchesi,[56,57] J. E. Brau,[58] T. Briant,[59] A. Brillet,[52] M. Brinkmann,[8] V. Brisson,[23] P. Brockill,[16] A. F. Brooks,[1] D. A. Brown,[35] D. D. Brown,[44] N. M. Brown,[10] C. C. Buchanan,[2] A. Buikema,[10] T. Bulik,[60] H. J. Bulten,[61,9] A. Buonanno,[29,62] D. Buskulic,[7] C. Buy,[30] R. L. Byer,[40] L. Cadonati,[63] G. Cagnoli,[64,65] C. Cahillane,[1] J. Calderón Bustillo,[66,63] T. Callister,[1] E. Calloni,[67,4] J. B. Camp,[68] K. C. Cannon,[69] J. Cao,[70] C. D. Capano,[8] E. Capocasa,[30] F. Carbognani,[34] S. Caride,[71] J. Casanueva Diaz,[23] C. Casentini,[25,13] S. Caudill,[16] M. Cavaglià,[21] F. Cavalier,[23] R. Cavalieri,[34] G. Cella,[19] C. B. Cepeda,[1] L. Cerboni Baiardi,[56,57] G. Cerretani,[18,19] E. Cesarini,[25,13] R. Chakraborty,[1] T. Chalermsongsak,[1] S. J. Chamberlin,[72] M. Chan,[36] S. Chao,[73] P. Charlton,[74] E. Chassande-Mottin,[30] H. Y. Chen,[75] Y. Chen,[76] C. Cheng,[73] A. Chincarini,[46]

A. Chiummo,[34] H. S. Cho,[77] M. Cho,[62] J. H. Chow,[20] N. Christensen,[78] Q. Chu,[50] S. Chua,[59] S. Chung,[50] G. Ciani,[5] F. Clara,[37] J. A. Clark,[63] F. Cleva,[52] E. Coccia,[25,12,13] P.-F. Cohadon,[59] A. Colla,[79,28] C. G. Collette,[80] L. Cominsky,[81] M. Constancio Jr.,[11] A. Conte,[79,28] L. Conti,[42] D. Cook,[37] T. R. Corbitt,[2] N. Cornish,[31] A. Corsi,[71] S. Cortese,[34] C. A. Costa,[11] M. W. Coughlin,[78] S. B. Coughlin,[82] J.-P. Coulon,[52] S. T. Countryman,[39] P. Couvares,[1] E. E. Cowan,[63] D. M. Coward,[50] M. J. Cowart,[6] D. C. Coyne,[1] R. Coyne,[71] K. Craig,[36] J. D. E. Creighton,[16] J. Cripe,[2] S. G. Crowder,[83] A. Cumming,[36] L. Cunningham,[36] E. Cuoco,[34] T. Dal Canton,[8] S. L. Danilishin,[36] S. D'Antonio,[13] K. Danzmann,[17,8] N. S. Darman,[84] V. Dattilo,[34] I. Dave,[47] H. P. Daveloza,[85] M. Davier,[23] G. S. Davies,[36] E. J. Daw,[86] R. Day,[34] D. DeBra,[40] G. Debreczeni,[38] J. Degallaix,[65] M. De Laurentis,[67,4] S. Deléglise,[59] W. Del Pozzo,[44] T. Denker,[8,17] T. Dent,[8] H. Dereli,[52] V. Dergachev,[1] R. De Rosa,[67,4] R. T. DeRosa,[6] R. DeSalvo,[87] C. Devine,[88] S. Dhurandhar,[14] M. C. Díaz,[85] L. Di Fiore,[4] M. Di Giovanni,[79,28] A. Di Lieto,[18,19] S. Di Pace,[79,28] I. Di Palma,[29,8] A. Di Virgilio,[19] G. Dojcinoski,[89] V. Dolique,[65] F. Donovan,[10] K. L. Dooley,[21] S. Doravari,[6,8] R. Douglas,[36] T. P. Downes,[16] M. Drago,[8,90,91] R. W. P. Drever,[1] J. C. Driggers,[37] Z. Du,[70] M. Ducrot,[7] S. E. Dwyer,[37] T. B. Edo,[86] M. C. Edwards,[78] A. Effler,[6] H.-B. Eggenstein,[8] P. Ehrens,[1] J. Eichholz,[5] S. S. Eikenberry,[5] W. Engels,[76] R. C. Essick,[10] Z. Etienne,[88] T. Etzel,[1] M. Evans,[10] T. M. Evans,[6] R. Everett,[72] M. Factourovich,[39] V. Fafone,[25,13,12] H. Fair,[35] S. Fairhurst,[92] X. Fan,[70] Q. Fang,[50] S. Farinon,[46] B. Farr,[75] W. M. Farr,[44] E. Fauchon-Jones,[92] M. Favata,[89] M. Fays,[92] H. Fehrmann,[8] M. M. Fejer,[40] I. Ferrante,[18,19] E. C. Ferreira,[11] F. Ferrini,[34] F. Fidecaro,[18,19] I. Fiori,[34] D. Fiorucci,[30] R. P. Fisher,[35] R. Flaminio,[65,93] M. Fletcher,[36] J.-D. Fournier,[52] S. Franco,[23] S. Frasca,[79,28] F. Frasconi,[19] Z. Frei,[53] A. Freise,[44] R. Frey,[58] V. Frey,[23] T. T. Fricke,[8] P. Fritschel,[10] V. V. Frolov,[6] P. Fulda,[5] M. Fyffe,[6] H. A. G. Gabbard,[21] S. M. Gaebel,[44] J. R. Gair,[94] L. Gammaitoni,[32,33] S. G. Gaonkar,[14] F. Garufi,[67,4] A. Gatto,[30] G. Gaur,[95,96] N. Gehrels,[68] G. Gemme,[46] B. Gendre,[52] E. Genin,[34] A. Gennai,[19] J. George,[47] L. Gergely,[97] V. Germain,[7] Archisman Ghosh,[15] S. Ghosh,[51,9] J. A. Giaime,[2,6] K. D. Giardina,[6] A. Giazotto,[19] K. Gill,[98] A. Glaefke,[36] E. Goetz,[99] R. Goetz,[5] L. Gondan,[53] G. González,[2] J. M. Gonzalez Castro,[18,19] A. Gopakumar,[100] N. A. Gordon,[36] M. L. Gorodetsky,[48] S. E. Gossan,[1] M. Gosselin,[34] R. Gouaty,[7] C. Graef,[36] P. B. Graff,[62] M. Granata,[65] A. Grant,[36] S. Gras,[10] C. Gray,[37] G. Greco,[56,57] A. C. Green,[44] P. Groot,[51] H. Grote,[8] S. Grunewald,[29] G. M. Guidi,[56,57] X. Guo,[70] A. Gupta,[14] M. K. Gupta,[96] K. E. Gushwa,[1] E. K. Gustafson,[1] R. Gustafson,[99] J. J. Hacker,[22] B. R. Hall,[55] E. D. Hall,[1] G. Hammond,[36] M. Haney,[100] M. M. Hanke,[8] J. Hanks,[37] C. Hanna,[72] M. D. Hannam,[92] J. Hanson,[6] T. Hardwick,[2] J. Harms,[56,57] G. M. Harry,[101] I. W. Harry,[29] M. J. Hart,[36] M. T. Hartman,[5] C.-J. Haster,[44] K. Haughian,[36] J. Healy,[102] A. Heidmann,[59] M. C. Heintze,[5,6] H. Heitmann,[52] P. Hello,[23] G. Hemming,[34] M. Hendry,[36] I. S. Heng,[36] J. Hennig,[36] A. W. Heptonstall,[1] M. Heurs,[8,17] S. Hild,[36] D. Hoak,[103] K. A. Hodge,[1] D. Hofman,[65] S. E. Hollitt,[104] K. Holt,[6] D. E. Holz,[75] P. Hopkins,[92] D. J. Hosken,[104] J. Hough,[36] E. A. Houston,[36] E. J. Howell,[50] Y. M. Hu,[36] S. Huang,[73] E. A. Huerta,[88,82] D. Huet,[23] B. Hughey,[98] S. Husa,[66] S. H. Huttner,[36] T. Huynh-Dinh,[6] A. Idrisy,[72] N. Indik,[8] D. R. Ingram,[37] R. Inta,[71] H. N. Isa,[36] J.-M. Isac,[59] M. Isi,[1] G. Islas,[22] T. Isogai,[10] B. R. Iyer,[15] K. Izumi,[37] T. Jacqmin,[59] H. Jang,[77] K. Jani,[63] P. Jaranowski,[105] S. Jawahar,[106] F. Jiménez-Forteza,[66] W. W. Johnson,[2] N. K. Johnson-McDaniel,[15] D. I. Jones,[26] R. Jones,[36] R. J. G. Jonker,[9] L. Ju,[50] Haris K,[107] C. V. Kalaghatgi,[24,92] V. Kalogera,[82] S. Kandhasamy,[21] G. Kang,[77] J. B. Kanner,[1] S. Karki,[58] M. Kasprzack,[2,23,34] E. Katsavounidis,[10] W. Katzman,[6] S. Kaufer,[17] T. Kaur,[50] K. Kawabe,[37] F. Kawazoe,[8,17] F. Kéfélian,[52] M. S. Kehl,[69] D. Keitel,[8,66] D. B. Kelley,[35] W. Kells,[1] R. Kennedy,[86] J. S. Key,[85] A. Khalaidovski,[8] F. Y. Khalili,[48] I. Khan,[12] S. Khan,[92] Z. Khan,[96] E. A. Khazanov,[108] N. Kijbunchoo,[37] C. Kim,[77] J. Kim,[109] K. Kim,[110] Nam-Gyu Kim,[77] Namjun Kim,[40] Y.-M. Kim,[109] E. J. King,[104] P. J. King,[37] D. L. Kinzel,[6] J. S. Kissel,[37] L. Kleybolte,[27] S. Klimenko,[5] S. M. Koehlenbeck,[8] K. Kokeyama,[2] S. Koley,[9] V. Kondrashov,[1] A. Kontos,[10] M. Korobko,[27] W. Z. Korth,[1] I. Kowalska,[60] D. B. Kozak,[1] V. Kringel,[8] B. Krishnan,[8] A. Królak,[111,112] C. Krueger,[17] G. Kuehn,[8] P. Kumar,[69] L. Kuo,[73] A. Kutynia,[111] B. D. Lackey,[35] M. Landry,[37] J. Lange,[102] B. Lantz,[40] P. D. Lasky,[113] A. Lazzarini,[1] C. Lazzaro,[63,42] P. Leaci,[29,79,28] S. Leavey,[36] E. O. Lebigot,[30,70] C. H. Lee,[109] H. K. Lee,[110] H. M. Lee,[114] K. Lee,[36] A. Lenon,[35] M. Leonardi,[90,91] J. R. Leong,[8] N. Leroy,[23] N. Letendre,[7] Y. Levin,[113] B. M. Levine,[37] T. G. F. Li,[1] A. Libson,[10] T. B. Littenberg,[115] N. A. Lockerbie,[106] J. Logue,[36] A. L. Lombardi,[103] L. T. London,[92] J. E. Lord,[35] M. Lorenzini,[12,13] V. Loriette,[116] M. Lormand,[6] G. Losurdo,[57] J. D. Lough,[8,17] C. O. Lousto,[102] G. Lovelace,[22] H. Lück,[17,8] A. P. Lundgren,[8] J. Luo,[78] R. Lynch,[10] Y. Ma,[50] T. MacDonald,[40] B. Machenschalk,[8] M. MacInnis,[10] D. M. Macleod,[2] F. Magaña-Sandoval,[35] R. M. Magee,[55] M. Mageswaran,[1] E. Majorana,[28] I. Maksimovic,[116] V. Malvezzi,[25,13] N. Man,[52] I. Mandel,[44] V. Mandic,[83] V. Mangano,[36] G. L. Mansell,[20] M. Manske,[16] M. Mantovani,[34] F. Marchesoni,[117,33] F. Marion,[7] S. Márka,[39] Z. Márka,[39] A. S. Markosyan,[40] E. Maros,[1] F. Martelli,[56,57] L. Martellini,[52] I. W. Martin,[36] R. M. Martin,[5] D. V. Martynov,[1] J. N. Marx,[1] K. Mason,[10] A. Masserot,[7] T. J. Massinger,[35] M. Masso-Reid,[36] F. Matichard,[10] L. Matone,[39] N. Mavalvala,[10]

N. Mazumder,[55] G. Mazzolo,[8] R. McCarthy,[37] D. E. McClelland,[20] S. McCormick,[6] S. C. McGuire,[118] G. McIntyre,[1] J. McIver,[1] D. J. McManus,[20] S. T. McWilliams,[88] D. Meacher,[72] G. D. Meadors,[29,8] J. Meidam,[9] A. Melatos,[84] G. Mendell,[37] D. Mendoza-Gandara,[8] R. A. Mercer,[16] E. Merilh,[37] M. Merzougui,[52] S. Meshkov,[1] C. Messenger,[36] C. Messick,[72] P. M. Meyers,[83] F. Mezzani,[28,79] H. Miao,[44] C. Michel,[65] H. Middleton,[44] E. E. Mikhailov,[119] L. Milano,[67,4] J. Miller,[10] M. Millhouse,[31] Y. Minenkov,[13] J. Ming,[29,8] S. Mirshekari,[120] C. Mishra,[15] S. Mitra,[14] V. P. Mitrofanov,[48] G. Mitselmakher,[5] R. Mittleman,[10] A. Moggi,[19] M. Mohan,[34] S. R. P. Mohapatra,[10] M. Montani,[56,57] B. C. Moore,[89] C. J. Moore,[121] D. Moraru,[37] G. Moreno,[37] S. R. Morriss,[85] K. Mossavi,[8] B. Mours,[7] C. M. Mow-Lowry,[44] C. L. Mueller,[5] G. Mueller,[5] A. W. Muir,[92] Arunava Mukherjee,[15] D. Mukherjee,[16] S. Mukherjee,[85] N. Mukund,[14] A. Mullavey,[6] J. Munch,[104] D. J. Murphy,[39] P. G. Murray,[36] A. Mytidis,[5] I. Nardecchia,[25,13] L. Naticchioni,[79,28] R. K. Nayak,[122] V. Necula,[5] K. Nedkova,[103] G. Nelemans,[51,9] M. Neri,[45,46] A. Neunzert,[99] G. Newton,[36] T. T. Nguyen,[20] A. B. Nielsen,[8] S. Nissanke,[51,9] A. Nitz,[8] F. Nocera,[34] D. Nolting,[6] M. E. Normandin,[85] L. K. Nuttall,[35] J. Oberling,[37] E. Ochsner,[16] J. O'Dell,[123] E. Oelker,[10] G. H. Ogin,[124] J. J. Oh,[125] S. H. Oh,[125] F. Ohme,[92] M. Oliver,[66] P. Oppermann,[8] Richard J. Oram,[6] B. O'Reilly,[6] R. O'Shaughnessy,[102] D. J. Ottaway,[104] R. S. Ottens,[5] H. Overmier,[6] B. J. Owen,[71] A. Pai,[107] S. A. Pai,[47] J. R. Palamos,[58] O. Palashov,[108] C. Palomba,[28] A. Pal-Singh,[27] H. Pan,[73] Y. Pan,[62] C. Pankow,[82] F. Pannarale,[92] B. C. Pant,[47] F. Paoletti,[34,19] A. Paoli,[34] M. A. Papa,[29,16,8] H. R. Paris,[40] W. Parker,[6] D. Pascucci,[36] A. Pasqualetti,[34] R. Passaquieti,[18,19] D. Passuello,[19] B. Patricelli,[18,19] Z. Patrick,[40] B. L. Pearlstone,[36] M. Pedraza,[1] R. Pedurand,[65] L. Pekowsky,[35] A. Pele,[6] S. Penn,[126] A. Perreca,[1] H. P. Pfeiffer,[69,29] M. Phelps,[36] O. Piccinni,[79,28] M. Pichot,[52] F. Piergiovanni,[56,57] V. Pierro,[87] G. Pillant,[34] L. Pinard,[65] I. M. Pinto,[87] M. Pitkin,[36] R. Poggiani,[18,19] P. Popolizio,[34] A. Post,[8] J. Powell,[36] J. Prasad,[14] V. Predoi,[92] S. S. Premachandra,[113] T. Prestegard,[83] L. R. Price,[1] M. Prijatelj,[34] M. Principe,[87] S. Privitera,[29] G. A. Prodi,[90,91] L. Prokhorov,[48] O. Puncken,[8] M. Punturo,[33] P. Puppo,[28] M. Pürrer,[29] H. Qi,[16] J. Qin,[50] V. Quetschke,[85] E. A. Quintero,[1] R. Quitzow-James,[58] F. J. Raab,[37] D. S. Rabeling,[20] H. Radkins,[37] P. Raffai,[53] S. Raja,[47] M. Rakhmanov,[85] P. Rapagnani,[79,28] V. Raymond,[29] M. Razzano,[18,19] V. Re,[25] J. Read,[22] C. M. Reed,[37] T. Regimbau,[52] L. Rei,[46] S. Reid,[49] D. H. Reitze,[1,5] H. Rew,[119] S. D. Reyes,[35] F. Ricci,[79,28] K. Riles,[99] N. A. Robertson,[1,36] R. Robie,[36] F. Robinet,[23] A. Rocchi,[13] L. Rolland,[7] J. G. Rollins,[1] V. J. Roma,[58] R. Romano,[3,4] G. Romanov,[119] J. H. Romie,[6] D. Rosińska,[127,43] C. Röver,[8] S. Rowan,[36] A. Rüdiger,[8] P. Ruggi,[34] K. Ryan,[37] S. Sachdev,[1] T. Sadecki,[37] L. Sadeghian,[16] L. Salconi,[34] M. Saleem,[107] F. Salemi,[8] A. Samajdar,[122] L. Sammut,[84,113] E. J. Sanchez,[1] V. Sandberg,[37] B. Sandeen,[82] J. R. Sanders,[99,35] B. Sassolas,[65] B. S. Sathyaprakash,[92] P. R. Saulson,[35] O. Sauter,[99] R. L. Savage,[37] A. Sawadsky,[17] P. Schale,[58] R. Schilling,[8,†] J. Schmidt,[8] P. Schmidt,[1,76] R. Schnabel,[27] R. M. S. Schofield,[58] A. Schönbeck,[27] E. Schreiber,[8] D. Schuette,[8,17] B. F. Schutz,[92,29] J. Scott,[36] S. M. Scott,[20] D. Sellers,[6] A. S. Sengupta,[95] D. Sentenac,[34] V. Sequino,[25,13] A. Sergeev,[108] G. Serna,[22] Y. Setyawati,[51,9] A. Sevigny,[37] D. A. Shaddock,[20] S. Shah,[51,9] M. S. Shahriar,[82] M. Shaltev,[8] Z. Shao,[1] B. Shapiro,[40] P. Shawhan,[62] A. Sheperd,[16] D. H. Shoemaker,[10] D. M. Shoemaker,[63] K. Siellez,[52,63] X. Siemens,[16] D. Sigg,[37] A. D. Silva,[11] D. Simakov,[8] A. Singer,[1] L. P. Singer,[68] A. Singh,[29,8] R. Singh,[2] A. Singhal,[12] A. M. Sintes,[66] B. J. J. Slagmolen,[20] J. R. Smith,[22] N. D. Smith,[1] R. J. E. Smith,[1] E. J. Son,[125] B. Sorazu,[36] F. Sorrentino,[46] T. Souradeep,[14] A. K. Srivastava,[96] A. Staley,[39] M. Steinke,[8] J. Steinlechner,[36] S. Steinlechner,[36] D. Steinmeyer,[8,17] B. C. Stephens,[16] S. P. Stevenson,[44] R. Stone,[85] K. A. Strain,[36] N. Straniero,[65] G. Stratta,[56,57] N. A. Strauss,[78] S. Strigin,[48] R. Sturani,[120] A. L. Stuver,[6] T. Z. Summerscales,[128] L. Sun,[84] P. J. Sutton,[92] B. L. Swinkels,[34] M. J. Szczepańczyk,[98] M. Tacca,[30] D. Talukder,[58] D. B. Tanner,[5] M. Tápai,[97] S. P. Tarabrin,[8] A. Taracchini,[29] R. Taylor,[1] T. Theeg,[8] M. P. Thirugnanasambandam,[1] E. G. Thomas,[44] M. Thomas,[6] P. Thomas,[37] K. A. Thorne,[6] K. S. Thorne,[76] E. Thrane,[113] S. Tiwari,[12] V. Tiwari,[92] K. V. Tokmakov,[106] C. Tomlinson,[86] M. Tonelli,[18,19] C. V. Torres,[85,‡] C. I. Torrie,[1] D. Töyrä,[44] F. Travasso,[32,33] G. Traylor,[6] D. Trifirò,[21] M. C. Tringali,[90,91] L. Trozzo,[129,19] M. Tse,[10] M. Turconi,[52] D. Tuyenbayev,[85] D. Ugolini,[130] C. S. Unnikrishnan,[100] A. L. Urban,[16] S. A. Usman,[35] H. Vahlbruch,[17] G. Vajente,[1] G. Valdes,[85] N. van Bakel,[9] M. van Beuzekom,[9] J. F. J. van den Brand,[61,9] C. Van Den Broeck,[9] D. C. Vander-Hyde,[35,22] L. van der Schaaf,[9] M. V. van der Sluys,[51] J. V. van Heijningen,[9] A. Vañó-Viñuales,[92] A. A. van Veggel,[36] M. Vardaro,[41,42] S. Vass,[1] M. Vasúth,[38] R. Vaulin,[10] A. Vecchio,[44] G. Vedovato,[42] J. Veitch,[44] P. J. Veitch,[104] K. Venkateswara,[131] D. Verkindt,[7] F. Vetrano,[56,57] A. Viceré,[56,57] S. Vinciguerra,[44] D. J. Vine,[49] J.-Y. Vinet,[52] S. Vitale,[10] T. Vo,[35] H. Vocca,[32,33] C. Vorvick,[37] D. Voss,[5] W. D. Vousden,[44] S. P. Vyatchanin,[48] A. R. Wade,[20] L. E. Wade,[132] M. Wade,[132] M. Walker,[2] L. Wallace,[1] S. Walsh,[16,8,29] G. Wang,[12] H. Wang,[44] M. Wang,[44] X. Wang,[70] Y. Wang,[50] R. L. Ward,[20] J. Warner,[37] M. Was,[7] B. Weaver,[37] L.-W. Wei,[52] M. Weinert,[8] A. J. Weinstein,[1] R. Weiss,[10] T. Welborn,[6] L. Wen,[50] P. Weßels,[8] T. Westphal,[8] K. Wette,[8] J. T. Whelan,[102,8] D. J. White,[86] B. F. Whiting,[5] R. D. Williams,[1] A. R. Williamson,[92] J. L. Willis,[133] B. Willke,[17,8] M. H. Wimmer,[8,17]

W. Winkler,[8] C. C. Wipf,[1] H. Wittel,[8,17] G. Woan,[36] J. Worden,[37] J. L. Wright,[36] G. Wu,[6] J. Yablon,[82] W. Yam,[10] H. Yamamoto,[1] C. C. Yancey,[62] M. J. Yap,[20] H. Yu,[10] M. Yvert,[7] A. Zadrożny,[111] L. Zangrando,[42] M. Zanolin,[98] J.-P. Zendri,[42] M. Zevin,[82] F. Zhang,[10] L. Zhang,[1] M. Zhang,[119] Y. Zhang,[102] C. Zhao,[50] M. Zhou,[82] Z. Zhou,[82] X. J. Zhu,[50] M. E. Zucker,[1,10] S. E. Zuraw,[103] and J. Zweizig[1]

(LIGO Scientific Collaboration and Virgo Collaboration)

M. Boyle,[134] B. Brügmann,[135] M. Campanelli,[102] M. Clark,[63] D. Hamberger,[76] L. E. Kidder,[134] M. Kinsey,[63] P. Laguna,[63] S. Ossokine,[29] M. A. Scheel,[76] B. Szilagyi,[76,136] S. Teukolsky,[134] and Y. Zlochower[102]

[1]LIGO, California Institute of Technology, Pasadena, California 91125, USA
[2]Louisiana State University, Baton Rouge, Louisiana 70803, USA
[3]Università di Salerno, Fisciano, I-84084 Salerno, Italy
[4]INFN, Sezione di Napoli, Complesso Universitario di Monte S.Angelo, I-80126 Napoli, Italy
[5]University of Florida, Gainesville, Florida 32611, USA
[6]LIGO Livingston Observatory, Livingston, Louisiana 70754, USA
[7]Laboratoire d'Annecy-le-Vieux de Physique des Particules (LAPP), Université Savoie Mont Blanc, CNRS/IN2P3, F-74941 Annecy-le-Vieux, France
[8]Albert-Einstein-Institut, Max-Planck-Institut für Gravitationsphysik, D-30167 Hannover, Germany
[9]Nikhef, Science Park, 1098 XG Amsterdam, Netherlands
[10]LIGO, Massachusetts Institute of Technology, Cambridge, Massachusetts 02139, USA
[11]Instituto Nacional de Pesquisas Espaciais, 12227-010 São José dos Campos, São Paulo, Brazil
[12]INFN, Gran Sasso Science Institute, I-67100 L'Aquila, Italy
[13]INFN, Sezione di Roma Tor Vergata, I-00133 Roma, Italy
[14]Inter-University Centre for Astronomy and Astrophysics, Pune 411007, India
[15]International Centre for Theoretical Sciences, Tata Institute of Fundamental Research, Bangalore 560012, India
[16]University of Wisconsin-Milwaukee, Milwaukee, Wisconsin 53201, USA
[17]Leibniz Universität Hannover, D-30167 Hannover, Germany
[18]Università di Pisa, I-56127 Pisa, Italy
[19]INFN, Sezione di Pisa, I-56127 Pisa, Italy
[20]Australian National University, Canberra, Australian Capital Territory 0200, Australia
[21]The University of Mississippi, University, Mississippi 38677, USA
[22]California State University Fullerton, Fullerton, California 92831, USA
[23]LAL, Université Paris-Sud, CNRS/IN2P3, Université Paris-Saclay, 91400 Orsay, France
[24]Chennai Mathematical Institute, Chennai 603103, India
[25]Università di Roma Tor Vergata, I-00133 Roma, Italy
[26]University of Southampton, Southampton SO17 1BJ, United Kingdom
[27]Universität Hamburg, D-22761 Hamburg, Germany
[28]INFN, Sezione di Roma, I-00185 Roma, Italy
[29]Albert-Einstein-Institut, Max-Planck-Institut für Gravitationsphysik, D-14476 Potsdam-Golm, Germany
[30]APC, AstroParticule et Cosmologie, Université Paris Diderot, CNRS/IN2P3, CEA/Irfu, Observatoire de Paris, Sorbonne Paris Cité, F-75205 Paris Cedex 13, France
[31]Montana State University, Bozeman, Montana 59717, USA
[32]Università di Perugia, I-06123 Perugia, Italy
[33]INFN, Sezione di Perugia, I-06123 Perugia, Italy
[34]European Gravitational Observatory (EGO), I-56021 Cascina, Pisa, Italy
[35]Syracuse University, Syracuse, New York 13244, USA
[36]SUPA, University of Glasgow, Glasgow G12 8QQ, United Kingdom
[37]LIGO Hanford Observatory, Richland, Washington 99352, USA
[38]Wigner RCP, RMKI, H-1121 Budapest, Konkoly Thege Miklós út 29-33, Hungary
[39]Columbia University, New York, New York 10027, USA
[40]Stanford University, Stanford, California 94305, USA
[41]Università di Padova, Dipartimento di Fisica e Astronomia, I-35131 Padova, Italy
[42]INFN, Sezione di Padova, I-35131 Padova, Italy
[43]CAMK-PAN, 00-716 Warsaw, Poland
[44]University of Birmingham, Birmingham B15 2TT, United Kingdom
[45]Università degli Studi di Genova, I-16146 Genova, Italy
[46]INFN, Sezione di Genova, I-16146 Genova, Italy

[47]RRCAT, Indore, Madhya Pradesh 452013, India
[48]Faculty of Physics, Lomonosov Moscow State University, Moscow 119991, Russia
[49]SUPA, University of the West of Scotland, Paisley PA1 2BE, United Kingdom
[50]University of Western Australia, Crawley, Western Australia 6009, Australia
[51]Department of Astrophysics/IMAPP, Radboud University Nijmegen, P.O. Box 9010, 6500 GL Nijmegen, Netherlands
[52]Artemis, Université Côte d'Azur, CNRS, Observatoire Côte d'Azur, CS 34229, Nice cedex 4, France
[53]MTA Eötvös University, "Lendulet" Astrophysics Research Group, Budapest 1117, Hungary
[54]Institut de Physique de Rennes, CNRS, Université de Rennes 1, F-35042 Rennes, France
[55]Washington State University, Pullman, Washington 99164, USA
[56]Università degli Studi di Urbino "Carlo Bo", I-61029 Urbino, Italy
[57]INFN, Sezione di Firenze, I-50019 Sesto Fiorentino, Firenze, Italy
[58]University of Oregon, Eugene, Oregon 97403, USA
[59]Laboratoire Kastler Brossel, UPMC-Sorbonne Universités, CNRS, ENS-PSL Research University, Collège de France, F-75005 Paris, France
[60]Astronomical Observatory Warsaw University, 00-478 Warsaw, Poland
[61]VU University Amsterdam, 1081 HV Amsterdam, Netherlands
[62]University of Maryland, College Park, Maryland 20742, USA
[63]Center for Relativistic Astrophysics and School of Physics, Georgia Institute of Technology, Atlanta, Georgia 30332, USA
[64]Institut Lumière Matière, Université de Lyon, Université Claude Bernard Lyon 1, UMR CNRS 5306, 69622 Villeurbanne, France
[65]Laboratoire des Matériaux Avancés (LMA), IN2P3/CNRS, Université de Lyon, F-69622 Villeurbanne, Lyon, France
[66]Universitat de les Illes Balears, IAC3—IEEC, E-07122 Palma de Mallorca, Spain
[67]Università di Napoli "Federico II", Complesso Universitario di Monte S.Angelo, I-80126 Napoli, Italy
[68]NASA/Goddard Space Flight Center, Greenbelt, Maryland 20771, USA
[69]Canadian Institute for Theoretical Astrophysics, University of Toronto, Toronto, Ontario M5S 3H8, Canada
[70]Tsinghua University, Beijing 100084, China
[71]Texas Tech University, Lubbock, Texas 79409, USA
[72]The Pennsylvania State University, University Park, Pennsylvania 16802, USA
[73]National Tsing Hua University, Hsinchu City, 30013 Taiwan, Republic of China
[74]Charles Sturt University, Wagga Wagga, New South Wales 2678, Australia
[75]University of Chicago, Chicago, Illinois 60637, USA
[76]Caltech CaRT, Pasadena, California 91125, USA
[77]Korea Institute of Science and Technology Information, Daejeon 305-806, Korea
[78]Carleton College, Northfield, Minnesota 55057, USA
[79]Università di Roma "La Sapienza", I-00185 Roma, Italy
[80]University of Brussels, Brussels 1050, Belgium
[81]Sonoma State University, Rohnert Park, California 94928, USA
[82]Northwestern University, Evanston, Illinois 60208, USA
[83]University of Minnesota, Minneapolis, Minneapolis 55455, USA
[84]The University of Melbourne, Parkville, Victoria 3010, Australia
[85]The University of Texas Rio Grande Valley, Brownsville, Texas 78520, USA
[86]The University of Sheffield, Sheffield S10 2TN, United Kingdom
[87]University of Sannio at Benevento, I-82100 Benevento, Italy and INFN, Sezione di Napoli, I-80100 Napoli, Italy
[88]West Virginia University, Morgantown, West Virginia 26506, USA
[89]Montclair State University, Montclair, New Jersey 07043, USA
[90]Università di Trento, Dipartimento di Fisica, I-38123 Povo, Trento, Italy
[91]INFN, Trento Institute for Fundamental Physics and Applications, I-38123 Povo, Trento, Italy
[92]Cardiff University, Cardiff CF24 3AA, United Kingdom
[93]National Astronomical Observatory of Japan, 2-21-1 Osawa, Mitaka, Tokyo 181-8588, Japan
[94]School of Mathematics, University of Edinburgh, Edinburgh EH9 3FD, United Kingdom
[95]Indian Institute of Technology, Gandhinagar Ahmedabad Gujarat 382424, India
[96]Institute for Plasma Research, Bhat, Gandhinagar 382428, India
[97]University of Szeged, Dóm tér 9, Szeged 6720, Hungary
[98]Embry-Riddle Aeronautical University, Prescott, Arizona 86301, USA
[99]University of Michigan, Ann Arbor, Michigan 48109, USA
[100]Tata Institute of Fundamental Research, Mumbai 400005, India
[101]American University, Washington, D.C. 20016, USA
[102]Rochester Institute of Technology, Rochester, New York 14623, USA
[103]University of Massachusetts-Amherst, Amherst, Massachusetts 01003, USA
[104]University of Adelaide, Adelaide, South Australia 5005, Australia
[105]University of Białystok, 15-424 Biał ystok, Poland

[106]SUPA, University of Strathclyde, Glasgow G1 1XQ, United Kingdom
[107]IISER-TVM, CET Campus, Trivandrum Kerala 695016, India
[108]Institute of Applied Physics, Nizhny Novgorod, 603950, Russia
[109]Pusan National University, Busan 609-735, Korea
[110]Hanyang University, Seoul 133-791, Korea
[111]NCBJ, 05-400 Świerk-Otwock, Poland
[112]IM-PAN, 00-956 Warsaw, Poland
[113]Monash University, Victoria 3800, Australia
[114]Seoul National University, Seoul 151-742, Korea
[115]University of Alabama in Huntsville, Huntsville, Alabama 35899, USA
[116]ESPCI, CNRS, F-75005 Paris, France
[117]Università di Camerino, Dipartimento di Fisica, I-62032 Camerino, Italy
[118]Southern University and A&M College, Baton Rouge, Louisiana 70813, USA
[119]College of William and Mary, Williamsburg, Virginia 23187, USA
[120]Instituto de Física Teórica, University Estadual Paulista/ICTP South American Institute for Fundamental Research, São Paulo, São Paulo 01140-070, Brazil
[121]University of Cambridge, Cambridge CB2 1TN, United Kingdom
[122]IISER-Kolkata, Mohanpur, West Bengal 741252, India
[123]Rutherford Appleton Laboratory, HSIC, Chilton, Didcot, Oxon OX11 0QX, United Kingdom
[124]Whitman College, 345 Boyer Avenue, Walla Walla, Washington 99362 USA
[125]National Institute for Mathematical Sciences, Daejeon 305-390, Korea
[126]Hobart and William Smith Colleges, Geneva, New York 14456, USA
[127]Janusz Gil Institute of Astronomy, University of Zielona Góra, 65-265 Zielona Góra, Poland
[128]Andrews University, Berrien Springs, Michigan 49104, USA
[129]Università di Siena, I-53100 Siena, Italy
[130]Trinity University, San Antonio, Texas 78212, USA
[131]University of Washington, Seattle, Washington 98195, USA
[132]Kenyon College, Gambier, Ohio 43022, USA
[133]Abilene Christian University, Abilene, Texas 79699, USA
[134]Cornell University, Ithaca, New York 14853, USA
[135]Theoretical Physics Institute, University of Jena, 07743 Jena, Germany
[136]Caltech JPL, Pasadena, California 91109, USA

[†]Deceased, May 2015.
[‡]Deceased, March 2015.